# Enhancing Laser Surface Texturing through Advanced Machine Learning Techniques

Christoph Zwahr[1*], Frederic Schell[1], Tobias Steege[1], Andrés Fabian Lasagni[1,2]

[1] Fraunhofer Institute for Material and Beam Technology IWS, Winterbergstraße 28, 01277 Dresden, Germany

[2] Institut für Fertigungstechnik, Technische Universität Dresden, George-Bähr-Str. 3c, 01069 Dresden, Germany

e-mails: christoph.zwahr@iws.fraunhofer.de; tobias.steege@iws.fraunhofer.de; frederic.schell@iws.fraunhofer.de; andres_fabian.lasagni@tu-dresden.de

*corresponding author

**Abstract**

Laser material processing has emerged as a versatile and indispensable tool in various industries, including manufacturing, healthcare, and materials science. However, the interaction of a lasers with surfaces is highly dependent on a large number of factors, including properties of the laser source such as pulse duration, wavelength and pulse form, as well as properties of the material such as surface roughness, heat capacity and thermal conductivity.  Therefore, the optimization of laser texturing processes in regards to specific target geometries while maintaining texture quality and process efficiency is a time consuming task that requires experienced operators with expert knowledge of the process and its components. The complex and nonlinear relationships between the various process, laser and material parameters and the resulting surface topography or functionality are challenging to model analytically. Therefore, the fabrication of large numbers of different parameter variations are typically required to enable empirical modeling and process optimization. Machine learning offers a promising approach to overcoming these challenges, particularly when the interrelations between process parameters are not well understood. It enables effective process optimization, surface property prediction, and automated monitoring—tasks that previously required expert knowledge. This chapter demonstrates the application of machine learning to Laser Surface Texturing techniques. Using algorithms such as neural networks and random forests, surface roughness can be predicted based on laser parameters and material data. This facilitates faster process optimization, reduces experimental effort, and enables predictive visualization—all while maintaining high accuracy.

# 1 Introduction

Laser Surface Texturing (LST) is a powerful technique that uses focused pulsed lasers to modify material surfaces at micro- and nanoscales, enabling control over physical, chemical, and biological properties. This process typically involves laser energy absorption by the material, inducing localized changes such as heating, melting, or ablation. LST methods include:

1. **Direct Laser Writing (DLW)**: Involves scanning a focused laser beam across a surface to create complex patterns. It is used for applications like microfluidic channels and microlens arrays.
2. **Direct Laser Interference Patterning (DLIP)**: Uses interference from multiple laser beams to create periodic surface patterns. DLIP is ideal for producing large, uniform textures, such as anti-reflective coatings for solar cells.
3. **Laser-Induced Periodic Surface Structures (LIPSS)**: Lasers can generate self-organized nanostructures, which can be for instance used for superhydrophobic surfaces and improved light absorption in photovoltaics.

LST offers non-contact, high-precision surface modification and can reach outstanding throughputs [1–3]. However, optimizing LST processes is challenging due to complex relationships between laser parameters and surface structures. Machine learning (ML) approaches are increasingly used to analyze data, predict surface outcomes, and optimize laser parameters, thus advancing the capabilities of LST.

## 1.1 AI-aided Surface Processing

The use of Artificial Intelligence (AI) in laser surface processing has gained significant attention due to its ability to optimize the precision, efficiency, and versatility of various machining processes. AI-assisted techniques, particularly those that integrate machine learning (ML) and deep learning (DL), enable predictive modeling, adaptive control, and real-time adjustments during laser machining. These advanced technologies enhance the precision of surface features such as roughness, depth, and width by analyzing large datasets and continuously improving the process. AI can predict machining outcomes and even optimize laser parameters for desired surface characteristics, reducing trial-and-error approaches and accelerating production. The integration of AI facilitates various aspects of laser surface processing, including parameter prediction, optimization of scanning paths, surface quality control, and functional property enhancement.

To better understand the approach to AI-assisted laser surface processing, a model developed by Ghosh et al. can be used [4]. This model encompasses a range of techniques for surface assessment, database preparation, modeling, evaluation, and deployment, as shown in **Fig. 19.1**. The model illustrates the key components of AI-assisted surface processing, highlighting the steps involved in optimizing the process through vision-based analysis, feature engineering, machine learning modeling, and iterative evaluation. The model provides a comprehensive framework for deploying AI techniques to predict and enhance surface topography, as well as important functional properties such as wettability, tribological behavior, and optical characteristics.

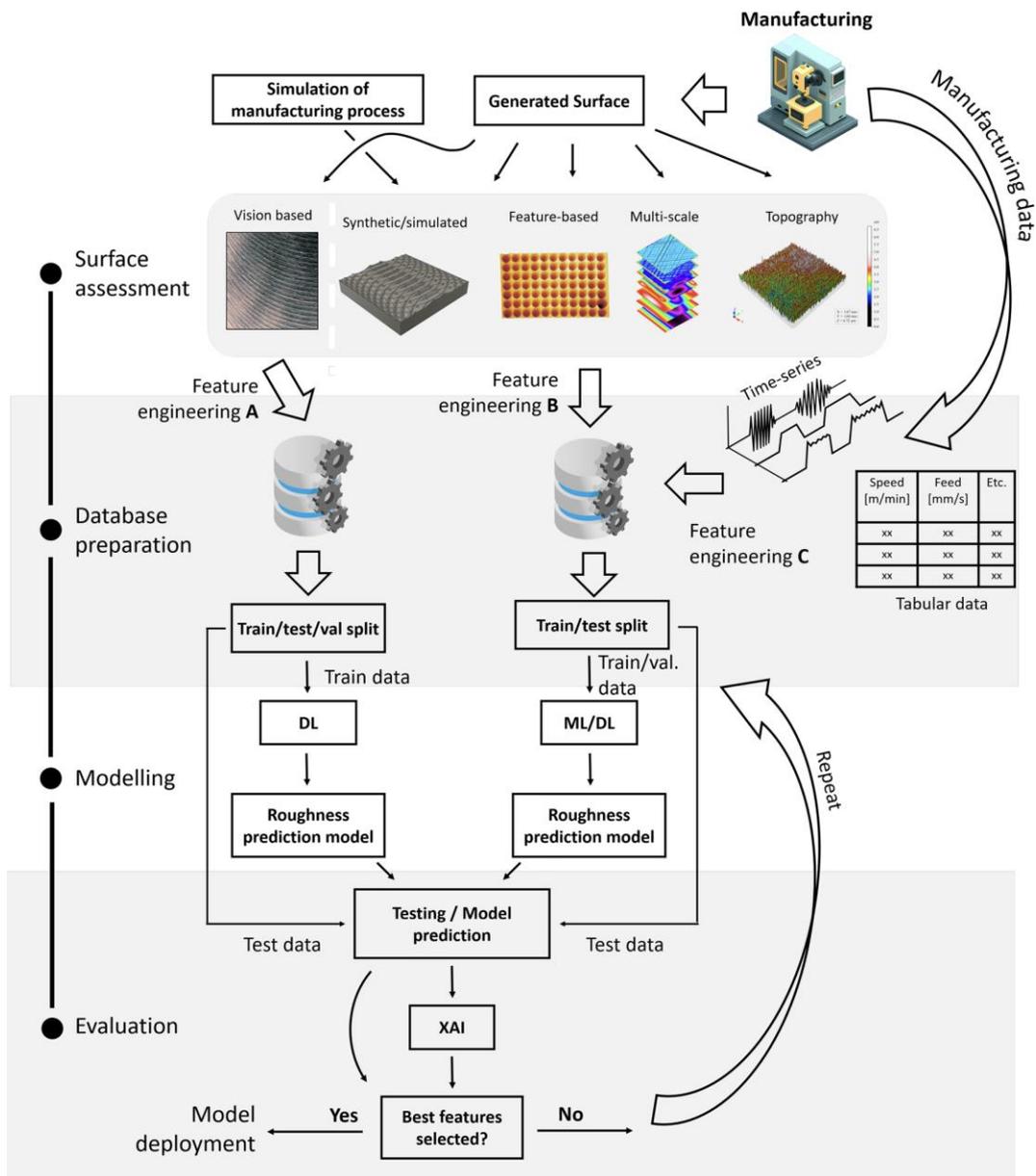

**Fig. 19.1:** Hierarchy of the typical processing steps for AI-aided surface roughness assessment from various manufacturing processes. The initial step consists of surfaces assessment by different methods, such as topographic measurement, vision-based assessment (cameras, microscopes), or synthetic topographies. In the following database preparation step, information for the first step as well as process parameters are aggregated and fed to the models in the third step. Finally, the models are evaluated and the process is repeated until a satisfactory model prediction capability is reached. Reprinted from Ghosh et al. [4], Towards AI driven surface roughness evaluation in manufacturing: a prospective study, Journal of Intelligent Manufacturing, 2024. Copyright 2024 under Creative Commons BY 4.0 license. Retrieved from https://doi.org/10.1007/s10845-024-02493-1.

The steps applied in the model are briefly explained as follows:

1. **Manufacturing Process Simulation**: The process starts with simulating the manufacturing process, generating a surface that is then assessed using various methods (such as vision-based, synthetic/simulated, feature-based, multi-scale, and topography).
2. **Feature Engineering**: These processes create various features from the generated surface data, which include different types of topography or time-series data from manufacturing inputs (like speed and feed).

3. **Database Preparation**: Data from different feature engineering approaches is prepared and split into train, validation, and test sets. This division ensures that the model can learn effectively and be evaluated on unseen data.
4. **Modeling**: The training data is then used for deep learning (DL) or machine learning (ML/DL) to predict surface roughness. Roughness prediction models are trained using this data.
5. **Evaluation**: After training, the model is tested using test data, and predictions are made. At this point, Explainable AI (XAI) is used to evaluate the model and select the most relevant features based on performance.
6. **Model Deployment**: Once the model passes the evaluation step, it is deployed, provided that the best features have been selected. If the best features are not identified during testing, the process may loop back for further refinement.

This process is iterated until the model reaches optimal performance, allowing the deployment of a robust roughness prediction model based on the manufacturing data and features derived from the simulated surface.

## 1.2 Analysis in Machine Learning Frameworks

Although the underlying techniques of AI have been known for decades, they only recently became viable to a large number of researchers as a consequence of the dramatic rise in computing power in the 21$^{st}$ century. Further contributing to the increased accessibility of machine-learning to a wide audience, high-performance machine-learning libraries written in low-level languages were released under open-source licenses and have received bindings to the high-level Python programming language, which is widely popular in the academic field due to its simple syntax, low learning curve and vast ecosystem of scientific computing libraries. Libraries such as Scikit-Learn [5] offer user friendly interfaces to machine learning algorithms such as Random Forest Regressors (RFR), Support Vector Machines (SVM) and Decision Trees (DT). More complex artificial neural networks (ANN) can be designed and applied using the prominent libraries TensorFlow [6] and PyTorch [7].

Machine learning differs fundamentally from traditional computing insofar as it treats the process as a black box. Traditional approaches require the implementation of specific algorithms, which often rely on an interplay of analytic modelling and known empirical relationships. This requires a fundamental understanding of the processes, which are to be modelled. However, processes such as laser surface texturing present a complex multivariate relationship between a large number of input parameters and output parameters. In LST, the properties of the laser source, such as the beam diameter, beam quality and shape, beam modes, wavelength, pulse duration, temporal pulse shape and other variables in conjunction with processing parameters such as the pulse overlap, scan speed, repetition rate and number of overpasses influence the final surface properties. The topographical properties on the other hand can be described by a large number of quantitative parameters, which capture different aspects of the topographies' geometry. Typically, surface roughness is quantified by parameters standardized in ISO 25178-2, compassing multiple parameter families that are related to various aspects of the surface topography such as the amplitude (Sa, Sq, …), the spatial properties (Sal, Str) and the functional properties derived from the bearing ratio curve (Sk, Spk, Svk, …).

Moreover, the laser process also influences the surface chemistry [8,9], contributing furthermore to the resulting physical interfacial properties in a way that is quite complex to model. Therefore, these processes lend themselves well to machine learning, where the relationship between input and output variables need not be understood, but is itself the subject of the machine learning output.

**Regression and Classification Algorithms**

Machine learning algorithms for regression and classification tasks can be broadly categorized by their architectural approaches. Two prominent methods that have shown significant success in predicting surface characteristics are ANNs and ensemble learning methods.

Artificial Neural Networks are computational models inspired by biological neural networks, mimicking the network of neurons in living organisms' nervous systems. They consist of interconnected artificial neurons that process information through weighted connections. The network architecture is organized in layers, with an input layer receiving data and an output layer producing results. Between these, there can be multiple hidden layers, with networks containing many hidden layers being referred to as deep learning network. The number of nodes in these layers must match the number of input and output parameters respectively. For surface prediction tasks, typical input parameters include laser frequency, power, and structuring speed, while the output often represents surface characteristics like average roughness. The network's accuracy is significantly influenced by its architecture, with recent trends favoring deep learning networks with hundreds of hidden layers, though this requires substantial training datasets to avoid overfitting [10].

Ensemble learning methods take a different approach by combining multiple machine learning models to produce more accurate and robust predictions than individual models. The fundamental principle is that a group of diverse models working together can outperform any single model by aggregating their predictions. Two primary approaches dominate ensemble learning: bagging and boosting. Random Forest exemplifies the bagging approach, combining multiple uncorrelated decision trees where each tree is trained on a random subset of the data and features. The number of decision trees can range from several hundred to thousands, with prediction accuracy generally improving as more trees are added. In contrast, boosting methods like AdaBoost train models sequentially, with each subsequent model focusing on correcting errors made by previous models. While ensemble methods often achieve high accuracy and can provide insights into feature importance, their computational complexity can limit real-time applications.

**Optimization Algorithms**

Optimization algorithms are computational methods designed to find the best possible solution from a set of available alternatives. In machine learning, these algorithms are particularly important for tuning model parameters and improving prediction accuracy. Two notable approaches are Genetic Algorithms (GA) and Grey Wolf Optimization (GWO).

Genetic Algorithms draw inspiration from natural evolution, employing mechanisms like selection, crossover, and mutation. The algorithm starts with a population of potential solutions, where each solution represents a set of model parameters. Through iterative generations, the fittest solutions are selected and combined to produce new, potentially better solutions. The fitness of each solution is evaluated using a predefined objective function, such as prediction accuracy. This evolutionary approach helps GA avoid local optima and explore a broader solution space, making it particularly effective for complex optimization problems with multiple parameters. The Grey Wolf Optimization algorithm, inspired by the social hierarchy and hunting behavior of grey wolves, has gained attention for its ability to balance exploration and exploitation in the search space. The algorithm categorizes the population into four groups (alpha, beta, delta, and omega) based on their fitness values. The three best solutions guide the rest of the search agents toward promising regions of the search space.

Both algorithms demonstrate the potential of nature-inspired optimization methods for improving machine learning model performance, though they differ in their computational requirements and application-specific effectiveness.

# 2 Process and Parameter Modeling

The optimization and control of LST processes require a deep understanding of how various parameters influence the resulting surface characteristics. Key factors such as laser power, pulse duration, scanning speed, and beam overlap must be precisely managed to achieve desired outcomes [11,12]. Accurately predicting these process parameters is essential to minimize reliance on time-consuming trial-and-error methods [13].

A major challenge in LST lies in forecasting the surface topography, including critical aspects like surface roughness, feature depth, and feature width [12,14,15]. Traditional approaches often involve labor-intensive experiments, making process optimization lengthy and resource-intensive [14]. Advanced modeling techniques are transforming this process by establishing clear correlations between input parameters and the resulting surface features, enabling more efficient and reliable predictions [16]. Ultimately, the success of LST depends not only on achieving specific topographical features but also on tailoring surface functionality [17]. Whether improving wettability, enhancing wear resistance, or optimizing optical properties, the ability to predict and control these functionalities is vital.

## 2.1 Prediction of Process Parameters

**Scanning Path Prediction**

Machine learning can be effectively applied to optimize scanning paths by minimizing processing time, reducing the number of pulses, and enhancing the efficient control of scanner mirrors and other system components. In study by Chen et al. a Genetic Algorithm (GA) was applied to optimize the scanning sequence in laser marking processes to minimize the total scanning path and reduce operating time [18]. By simulating the scanning process, the algorithm determined the shortest route for the laser scanner to move between objects, similar to solving a Traveling Salesman Problem. The GA used Real-Coded Genetic Algorithm (RCGA) with Dynamic Variable Length Two Point Crossover (DVL-2PC) to generate and evaluate optimized solutions, improving scanning. The optimization achieved through the GA significantly reduced the total scanning time and path length and thereby also laser-on-time which reduced energy consuption and extends laser diode´s lifespan.

In a study by Zhang et al. [13] an Adaptive Discrete Grey Wolf Optimizer (A-D-GWO) to optimize scan paths in laser machining was employed, with the process modeled as a Traveling Salesman Problem (TSP). Each wolf represented a potential path solution, and the algorithm iteratively refined the paths by leveraging exchange, shift, and 2-opt operations. Hamming distances between solutions were used to guide the optimization process, ensuring closer alignment with the optimal solution. In **Fig. 19.2**, the drilling paths were represented as a graph, with vertices corresponding to holes and edges weighted by jump times. The optimization process included three main steps:

1. Initialization: The hamming distance was calculated, and mismatched elements in the sequence were identified.
2. Exploration and Exploitation: Random shifts introduced diversity, while targeted exchanges reduced mismatches.
3. Refinement: The 2-opt operation shortened paths locally, refining solutions further.

The experimental results demonstrated significant time reductions in processing files containing 90, 286, and 649 holes, with overall time savings of 38.9%, 53.6%, and 55.5%, respectively, compared to the initial configurations.

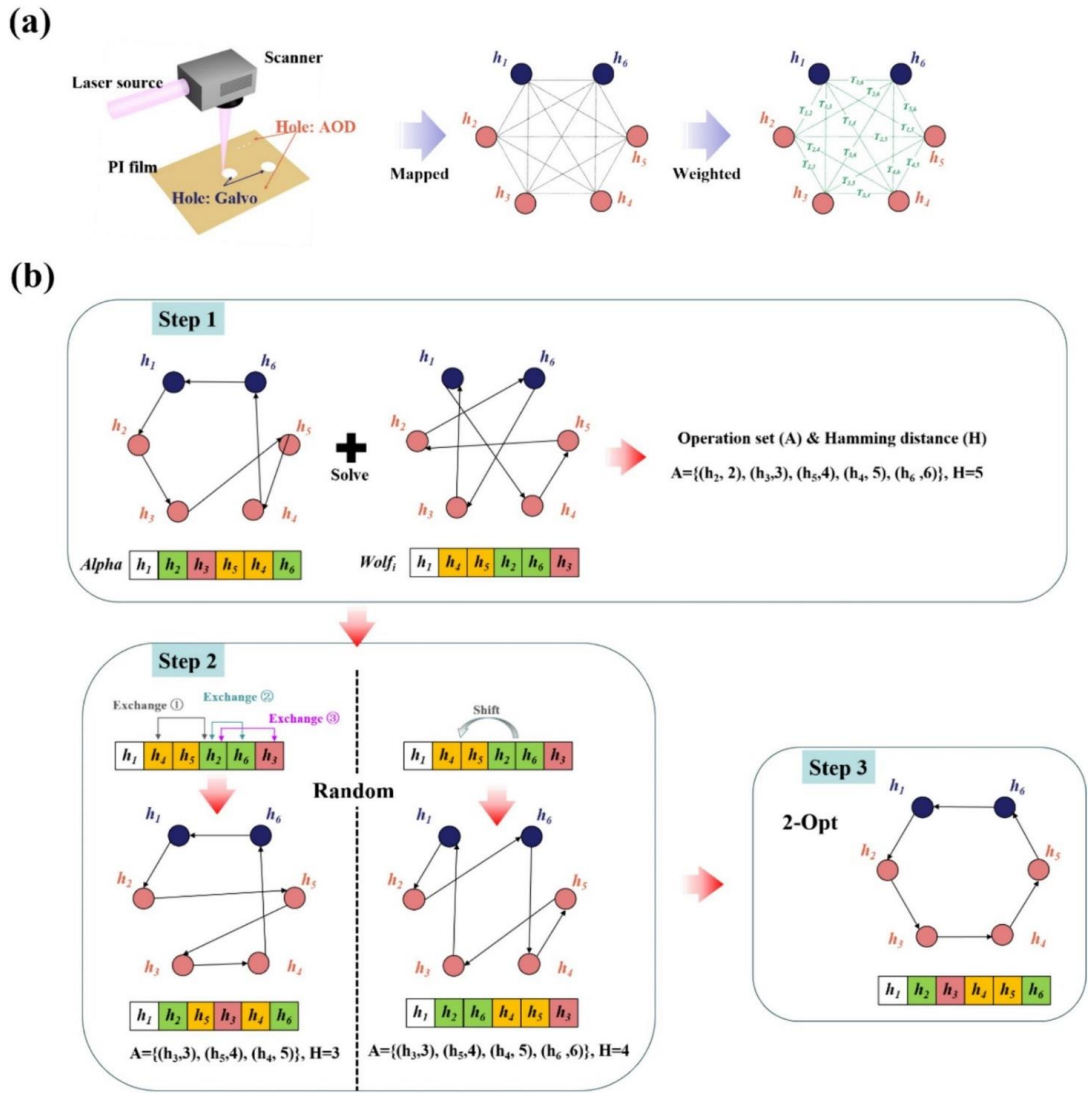

**Fig. 19.2:** Illustration of the methodology of the proposed path optimization process using the Adaptive Discrete Grey Wolf Optimizer (A-D-GWO). The figure is divided into two main components: (a) Graph representation of paths of drilling, containing vertices and edges. Each vertex corresponds to a micro-hole that needs to be machined and each edge represent potential jump paths between holes. (b) Discrete grey wolf optimization solver: The A-D-GWO workflow models the optimization process, inspired by the hunting behavior of grey wolves. Each "wolf" represents a feasible path solution, and the optimization is carried out through iterative refinements guided by the best solutions. Reprinted from [13], Zhang, T., Hu, H., Liang, Y., Liu, X., Rong, Y. & Wu, C. et al. (2024) A novel path planning approach to minimize machining time in laser machining of irregular micro-holes using adaptive discrete grey wolf optimizer. Computers & Industrial Engineering, 193, 110320, Copyright 2024, with permission from Elsevier.

**Process Optimization**

AI has emerged as a transformative tool in optimizing laser machining processes, addressing challenges like parameter selection, prediction accuracy, and efficiency. Machine learning models such as Artificial Neural Networks (ANNs), XGBoost, and hybrid approaches like Genetic Algorithm-tuned ANFIS have been applied to predict critical machining parameters, including depth, width, and kerf characteristics, with high precision [19–21]. AI techniques also enable inverse modeling to propose input parameters for desired outcomes, reducing experimental time and errors significantly. Advanced models like Grey Wolf Optimization-enhanced Backpropagation Neural Networks (GWO-BPNN) further improve prediction accuracy and adaptability [22]. Additionally, automated data acquisition systems integrated with AI facilitate real-time optimization and quality control, paving the way for smart and adaptive laser processing systems [23].

Important for the optimization of LST processes is the right choice of ML algorithms. Usually, different algorithms are tested and analyzed in terms of their accuracy regarding the target parameter. Moles et al. [16]focused on leveraging ML techniques to optimize the selection of femtosecond laser parameters for surface texturing, which aimed to enhance tribological performance. To address this, six ML models—Decision Trees, Random Forest, Gaussian Process, Support Vector Machines, K-Nearest Neighbors, and Artificial Neural Networks—were compared to predict the depth of laser textures, specifically lines and dimples. The data were preprocessed to remove inconsistencies and engineered to extract relevant features such as laser fluence, pulse energy, and scanning speed. Each model underwent hyperparameter optimization using the Optuna AutoML tool to maximize predictive accuracy. Validation was conducted through 3-fold cross-validation and the evaluation of $R^2$ scores on a test dataset. Among the models, ANNs were found to outperform the others, achieving the highest accuracy across all datasets. The entire process is illustrated in **Fig. 19.3a**, which depicts the ML pipeline. It also includes the integration of an inverse modelling approach to prescribe input laser parameters based on desired depths. **Fig. 19.3**b shows the desired groove and dimple textures to improve tribological performance.

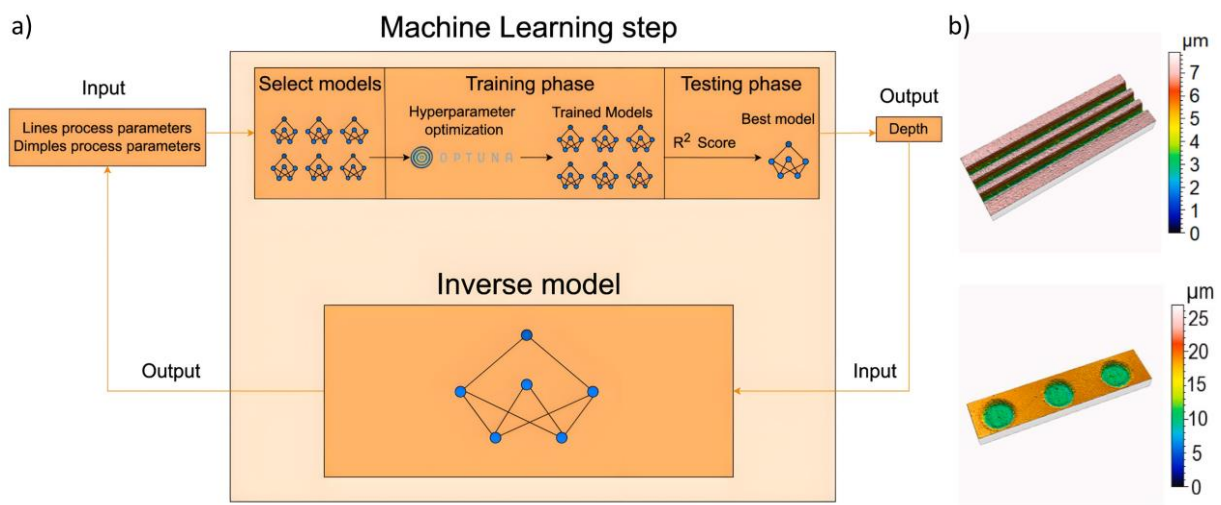

**Fig. 19.3:** (a) The Machine Learning pipeline used in the study, showcasing the sequential steps: data cleaning to remove inconsistencies, feature engineering to select relevant variables, model selection and hyperparameter optimization for improved accuracy, validation using test datasets to assess performance, and inverse modelling to prescribe optimal laser parameters based on target depths. (b) Detail of three femtosecond textured lines and dimples. Reprinted from [16], Moles, L., Llavori, I., Aginagalde, A., Echegaray, G., Bruneel, D. & Boto, F. et al. (2024) On the use of machine learning for predicting femtosecond laser grooves in tribological applications. Tribology International, 200, 110067, Copyright 2024, with permission from Elsevier.

To determine laser parameters, two inverse modelling approaches were proposed for achieving target depths. In the first approach, the trained ANN model was used to identify parameters from the training dataset that corresponded to the closest predicted depth. While this approach reduced errors to some extent, discrepancies in parameter estimation were observed due to the reliance on existing data points. To improve accuracy, a second approach based on a Genetic Algorithm was implemented. This algorithm iteratively optimized parameter combinations, significantly reducing prediction errors and improving alignment with the desired target depths. The analysis revealed that dimples' depths were predicted more accurately than those of lines, a trend that was consistent across all models. Statistical tests, including the Friedman and Nemenyi tests, confirmed the superior performance of ANNs, particularly in handling data variability and producing robust predictions. [16]

The optimization of laser processing parameters has been extensively studied using various advanced algorithms and hybrid models to improve accuracy and efficiency. In one study, the kerf width in laser machining of titanium alloys was optimized. A Genetic Algorithm-tuned Adaptive Neuro-Fuzzy Inference System (GA-ANFIS) was applied, and it was found to outperform traditional ANFIS models in prediction accuracy, demonstrating the efficacy of hybrid models for precise parameter optimization in laser machining. [20]

In a study by Vo et al., a smart system for optimizing laser processing parameters was developed to address the challenges posed by large parameter spaces and material variability [23]. An Automated Data Acquisition System was integrated with an ANN, where laser-induced damage areas were measured and used for training and prediction. The ANN accurately modelled and optimized laser processing parameters, efficiently handling large datasets and proving suitable for tasks such as quality control and failure prediction. **Fig. 19.4** illustrates the optimization of the efficiency, defined as the ratio of the crater area to the total energy used to generate the crater. Total energy was calculated as the product of the pulse number (N) and the pulse energy. The relationship between efficiency and total energy was modelled for various pulse numbers using predictions from the ANN and confirmed by experiments showing good matching between prediction and validation.

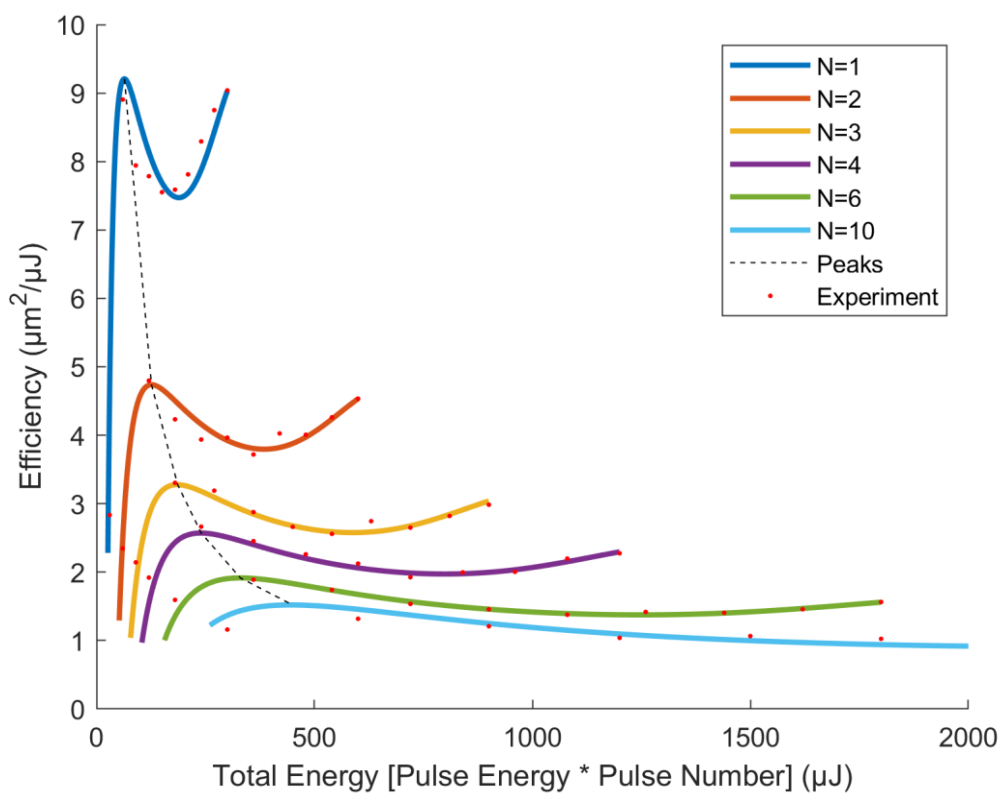

**Fig. 19.4:** Prediction and validation of the efficiency, defined as the ratio of the crater area to the total energy used to generate the crater, as a function of the total energy (product of pulse energy and pulse number N) for different pulse numbers. Reprinted with permission from [23], Vo, C., Zhou, B. & Yu, X. (2021) Optimization of laser processing parameters through automated data acquisition and artificial neural networks. Journal of Laser Applications, 33(4). Copyright 2021, Laser Institute of America.

The exploration of advanced algorithms in laser processing optimization highlighted two promising approaches. GA-ANFIS is best suited for applications requiring precise optimization in well-defined parameter spaces, combining the interpretability of fuzzy logic with the optimization capabilities of genetic algorithms. ANN with automated data acquisition proves the most versatile for real-time, scalable applications, effectively managing dynamic datasets and material variability while enabling broader tasks like quality control and adaptive processing. For static or targeted optimization, hybrid models like GA-ANFIS and GWO-BPNN are most effective, while ANN-based systems integrated with automated data acquisition are ideal for dynamic and large-scale adaptive applications.

**Predictive Visualization of Laser Surface Texturing**

Predictive visualization represents a cutting-edge approach recently adopted for LST processes. The primary goal is to forecast the outcomes of a machining process by generating visual representations of the surface before actual processing. These visualizations, which can take the form of topographical or microscopical images, are valuable for optimizing process parameters and enhancing efficiency [24–27]. The algorithms used for this purpose predominantly include Deep Neural Networks, which function as generative models, similar to popular tools like DALL-E . These networks synthesize realistic images based on input parameters, providing insights into the anticipated surface features and their tribological properties.

B. Mills et al. utilized a Generative Adversarial Network (GAN) to predict laser-ablated surface profiles resulting from three sequential pulse irradiations on an electroless nickel mirror. In their studies, the neural network accepted input in the form of three beam profiles modulated by a digital mirror device and produced output images of the laser-ablated surface [25–27]. To produce the results in **Fig. 19.5**, a neural network was trained using a dataset of 5,000 images, representing laser-machined features created from various spatial intensity patterns. Experimental 3D surface profiles of these features were collected and used for comparison. Two NN architectures, a Convolutional Neural Network (CNN) and a Conditional Adversarial Network (CAN), were evaluated. Both were trained over 100 epochs, with predictions analyzed at different training stages. The CAN outperformed the CNN, generating more realistic features such as burrs, non-flat bottoms, and scattered debris.

The trained NN was used to predict machining profiles for unseen patterns, including regular and inverted "X" shapes. These predictions were compared to experimental profiles, showing high accuracy in depth, width, and cross-sectional shape. Even challenging features, such as burr height and debris distribution, were replicated effectively. Minor deviations were attributed to noise in the experimental setup and the NN's deterministic nature. [25]

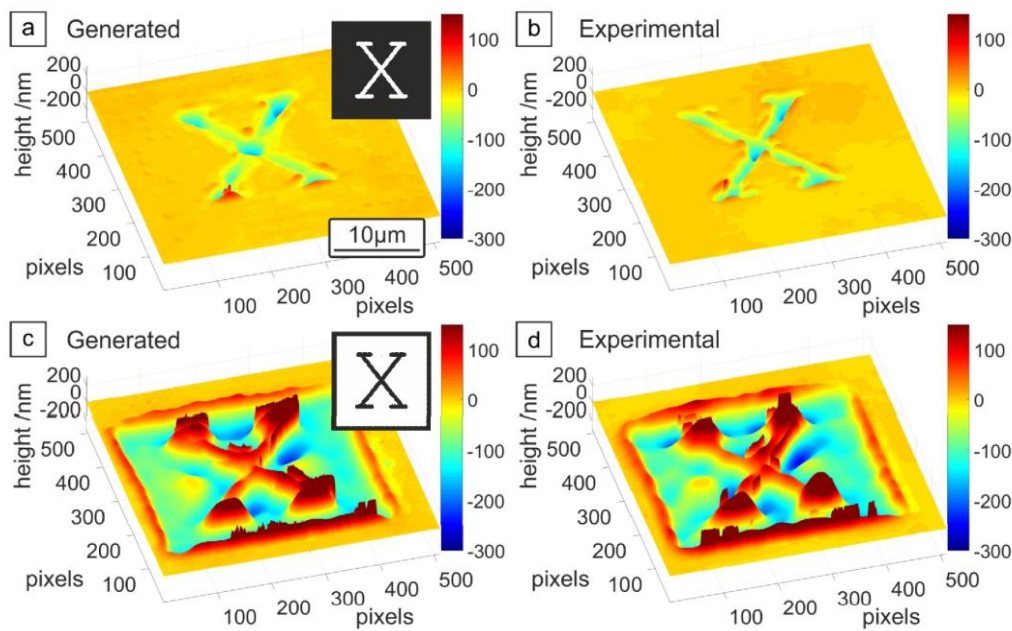

**Fig. 19.5:** Comparison of generated and experimentally measured images for spatial intensity patterns shaped into the letter 'X' in (a) and (b) respectively, and an inverted intensity 'X' in (c) and (d). Clear similarities are observed for both cases, in particular the depth, width and profile of the laser machined features, along with analytically difficult to predict features, such as burr height. Reprinted from [25], Heath, D.J., Grant-Jacob, J.A., Xie, Y., Mackay, B.S., Baker, J.A.G. & Eason, R.W. et al. (2018) Machine learning for 3D simulated visualization of laser machining. *Optics express*, 26(17), 21574. Copyright 2018 under Creative Commons BY 4.0 license. Retrieved from https://doi.org/10.1364/OE.26.021574.

The prediction of laser ablated geometries was brought one step further for multi-pulse ablation including scanning trajectories by Tani et al. [24]. They employed Deep Neural Networks (DNNs) to model the changes in surface morphology caused by single and multiple-pulse laser ablation. The architecture comprised four interconnected neural networks, which extracted material properties from local surface morphologies, computed nonlinear and nonlocal light-matter interactions, and generated two-dimensional depth profile changes induced by laser irradiation. The simulator used depth profiles and laser parameters as inputs to produce feature vectors representing material properties, including accumulated changes from preceding pulses. In the case of multiple-pulse irradiation, the simulator accurately replicated experimental depth profiles along complex laser trajectories, including circular and star-shaped patterns, as shown in **Fig. 19.6**. Quantitative comparisons revealed a strong agreement between the simulated and experimental data across various pulse energies, validating the simulator's ability to reproduce both average depths and surface roughness without relying on adjustable parameters.

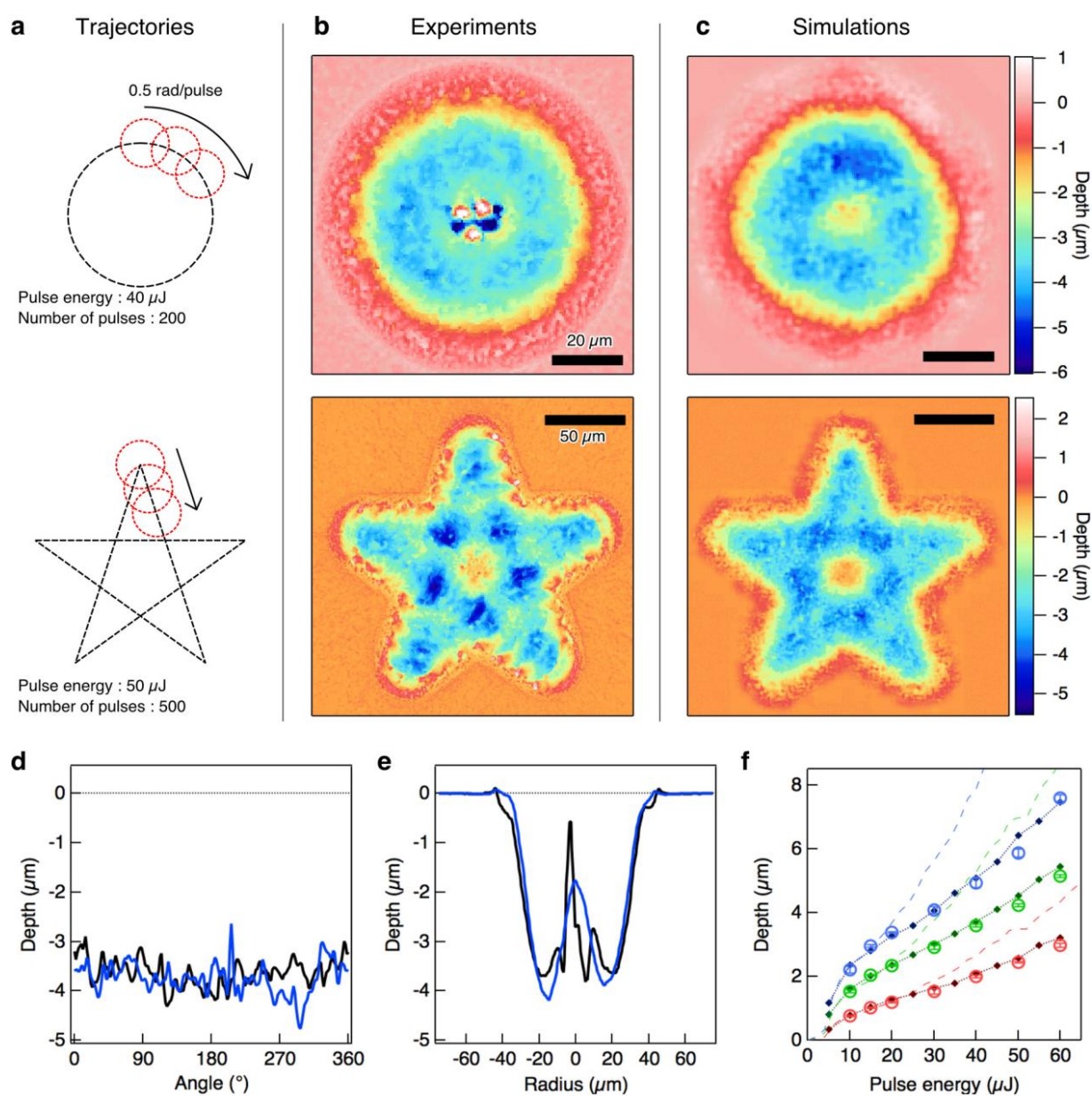

**Fig. 19.6:** Comparison of experiments with simulations for multi-shot irradiation. (a) Trajectories for multi-shot irradiation. (b,c) Final depth profiles after irradiation obtained by experiments and simulations, respectively. (d,e) One-dimensional depth profiles on circular irradiation from the simulation (blue lines) and the experiment

(black lines). (d) Depth profiles along the circular trajectory. (e) Depth profiles along the radial cross-section. (f) Averaged ablated depth along the circular trajectory as a function of the pulse energy from the simulation (open circles) and the experiment (filled diamonds). Intact initial surfaces were irradiated using 100 (red), 200 (green), or 300 (blue) pulses. Dashed lines represent the calculation reproduced from the arithmetic mean of the ablated depth over the training datasets for each of the pulse energies. A silicon substrate was used as the test target. Reprinted from [24], Tani, S. & Kobayashi, Y. (2022) Ultrafast laser ablation simulator using deep neural networks. *Scientific Reports*, 12(1), 5837. Copyright 2021 under Creative Commons BY 4.0 license. Retrieved from https://doi.org/10.1038/s41598-022-09870-x.

## 2.2 Prediction of Topographical Parameters

Parameter prediction in laser surface texturing can be realized through four distinct approaches [28]:

- Analytical Models: Based on machining theory and physical principles
- Experimental Models: Examining the influence of various factors through direct testing
- Design of Experiments Models: Systematic approach to parameter relationship analysis
- Artificial Intelligence (AI)-based Models: Using machine learning techniques for prediction

Traditional analytical and experimental models, while widely adopted in laser-based manufacturing, face several limitations. These approaches typically rely on statistical regression or analysis of variance and require constant adjustment for environmental, mechanical, and material influences. While they can provide useful insights, they often struggle to capture the complex nonlinear relationships between process parameters and resulting surface characteristics. Moreover, purely analytical modeling of laser machining processes is highly complex, requiring numerous machine-specific and material-related parameters to be determined empirically. Notably, no comprehensive analytical model currently exists for predicting average surface roughness (Sa) in laser texturing applications. In contrast, AI-based models have emerged as a preferred approach for parameter prediction in laser surface texturing. These models employ various techniques including:

- Artificial Neural Networks (ANNs): Well-established for handling nonlinear problems
- Random Forests (RFs): Known for their flexibility and effectiveness in both classification and regression tasks
- Genetic Algorithms (GA): Useful for generating high-quality solutions in optimization problems

**Surface Roughness Prediction**

Surface roughness is one of the most critical parameters in laser surface texturing, as it directly influences functional properties like wettability [29,30], tribological behavior [31,32], and anti-icing characteristics [33]. Studies have shown that machine learning approaches can effectively predict surface roughness based on laser processing parameters. For instance, Tani and Kobayashi [24] demonstrated that Deep Neural Networks can successfully predict surface modifications in ultrafast laser ablation processes, while Steege et al. [34] showed that both artificial neural networks and random forests can achieve high prediction accuracies for surface roughness in direct laser writing applications.

In the work of Steege et al. [34], both ANN and random forest (RF) approaches demonstrated strong predictive capabilities for laser surface texturing. Using a 100 ns pulsed ytterbium fiber laser (1064 nm wavelength, 100 μm spot diameter, 3 mJ pulse energy at 10 kHz), surface with random patterns were fabricated. Initial studies focused on understanding how different laser processing parameters influenced the resulting surface morphology. As shown in **Fig. 19.7**, systematic parameter studies on stainless steel (316L) and Stavax (AISI 420) surfaces revealed distinct correlations between laser processing parameters and resulting surface morphologies. The color matrices represent different roughness values that can be achieved on the material, with corresponding SEM images showing the resulting surface structures. These studies formed the basis for training the machine learning models, capturing the complex relationships between processing parameters and surface roughness.

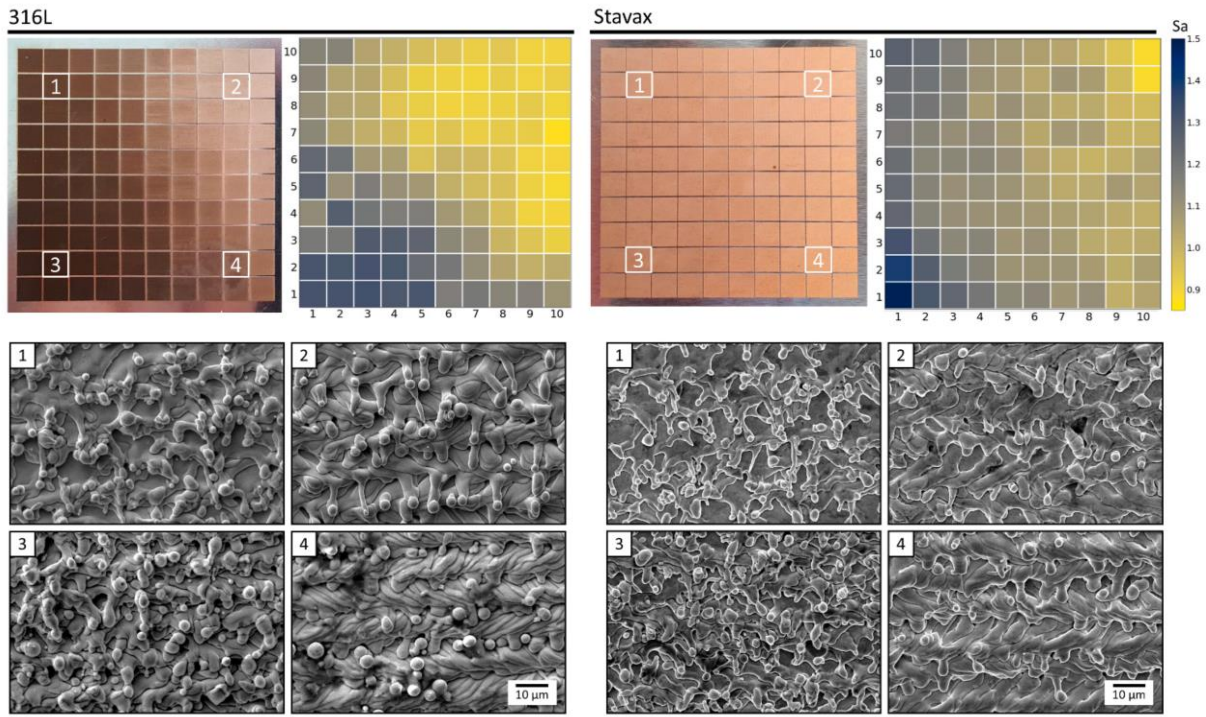

**Fig. 19.7:** Exemplary laser treated surface and resulting roughness (Sa) values for constant laser power of 18 W and frequencies (repetition rate) from 110 kHz to 200 kHz (bottom–top) and scanning speed from 1.1 m/s to 2.0 m/s (left to right) for the materials 316L (left) and Stavax (right). SEM images of the textured surface for both materials and four different process parameters. Reprinted from [34], Steege, T., Bernard, G., Darm, P., Kunze, T. & Lasagni, A.F. (2023) Prediction of Surface Roughness in Functional Laser Surface Texturing Utilizing Machine Learning. *Photonics*, 10(4), 361. Copyright 2023 under Creative Commons BY 4.0 license. Retrieved from https://doi.org/10.3390/photonics10040361.

Therefore, a comprehensive experimental dataset was created covering the complete parameter space of the laser system. The experiments involved varying three key processing parameters: laser pulse frequency (10-200 kHz), scanning speed (100-2500 mm/s), and laser power (15-30 W). For each material, eight parameter matrices containing 100 individual textures each were generated, resulting in 800 unique parameter combinations. Each textured region was 3 × 3 mm in size with a fixed hatch spacing of 50 μm. Surface roughness measurements were performed using white light interferometry with a 50× objective, providing both lateral (340 nm) and vertical (4 nm) resolution. The dataset was split into 80% training and 20% validation sets, with additional derived parameters like pulse overlap, pulse energy, and cumulative fluence dose included to improve model accuracy.

Both artificial neural networks ANN and RF approaches demonstrated strong predictive capabilities, as illustrated in **Fig. 19.8**. The comparison between predicted and measured surface roughness values shows high correlation, particularly in the range around 1 μm where most experimental data points were concentrated. For stainless steel, approximately 79.6% of the training data fell within this range, while for Stavax it was about 81.1%. The models achieved correlation factors ($R^2$) of over 0.9 for Stavax and approximately 0.8 for stainless steel, with mean absolute errors as low as 0.047 μm for Stavax.

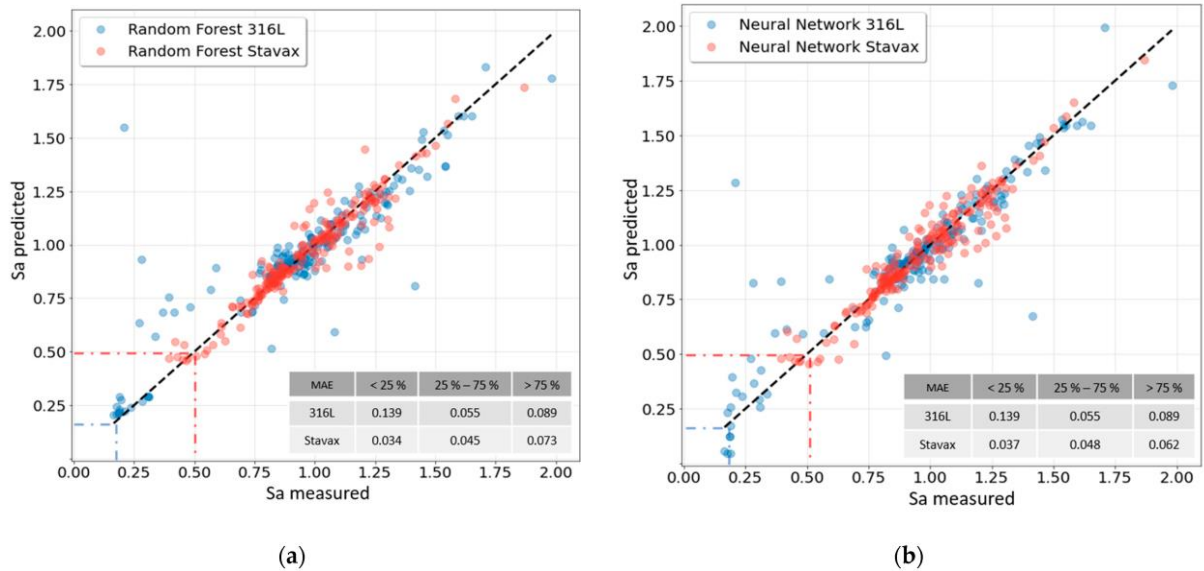

**Fig. 19.8:** Comparison of predicted and measured surface roughness Sa for RF (a) and ANN (b) for both materials: 316l (blue) and Stavax (red). The initial roughness for both samples is shown as dotted lines. The maximum absolute error (MAE) for the three ranges: lower 25%, upper 25%, and in between, is shown in the corresponding table. Reprinted from [34], Steege, T., Bernard, G., Darm, P., Kunze, T. & Lasagni, A.F. (2023) Prediction of Surface Roughness in Functional Laser Surface Texturing Utilizing Machine Learning. *Photonics*, 10(4), 361. Copyright 2023 under Creative Commons BY 4.0 license. Retrieved from https://doi.org/10.3390/photonics10040361.

Interestingly, surface roughness values below 0.5 µm were predicted more accurately for Stavax compared to stainless steel, which can be attributed to the higher initial surface roughness of Stavax leading to a polishing effect during laser processing with lower pulse energies. The models also revealed that surface roughness prediction accuracy varies between materials, with Stavax showing better overall prediction performance, likely due to its more consistent material properties resulting from the electro-slag remelting (ESR) process used in its production.

A study by Teixidor et al. [21] examined the efficacy of different machine learning approaches for predicting laser micromachining outcomes. Their research revealed that ANNs and decision trees consistently outperformed other methods like k-Nearest Neighbors and linear regression across multiple output parameters. Specifically, ANNs proved more effective for modeling channel width dimensions (achieving $R^2$ of ~0.61), while decision trees displayed better performance when predicting surface roughness ($R^2$ of ~0.60). For depth and material removal rate predictions, both techniques demonstrated similar high accuracy levels ($R^2 > 0.80$). The study highlighted that the modeling complexity varies significantly between different output parameters - with surface roughness being particularly challenging to predict accurately due to its inherent noisiness and dependence on multiple variables. This suggests that selecting the appropriate machine learning technique depends heavily on which specific output parameter is most critical for the intended application.

Based on this research by Moles et al., [35] the established benefits of machine learning for laser parameter optimization were validated specifically for femtosecond laser texturing. Using a Yb:YAG femtosecond laser system (280 fs pulse duration, 1030 nm wavelength), the researchers conducted experiments on steel and TiN/TiCN-coated surfaces, creating both line and dimple patterns. The experimental design systematically varied pulse repetition rate (50-166 kHz), pulse energy (1.57-12.41 µJ), and scanning speed (75-400 mm/s). Comparing six machine learning algorithms, they found that Artificial Neural Networks delivered the most accurate predictions of texture depth, with $R^2$ values up to 0.99 for dimples and 0.89 for lines. They also demonstrated that an evolutionary-based inverse model could successfully prescribe laser parameters to achieve target depths with mean errors below 5%. This shows that machine learning approaches previously proven successful for other laser processes can be effectively adapted for different laser based texturing approaches, e.g. femtosecond laser texturing.

Petit et al. similarly applied XGBoost and neural network approaches for predicting femtosecond laser micromachining parameters on 316L stainless steel [19]. Their study utilized an experimental dataset of 14,000 samples, with laser parameters including frequency (100-500 kHz), scanning speed (50-2000 mm/s), power (30-80% of 13.6 W), and spot diameter (14 and 26 μm). The XGBoost model achieved superior performance compared to the neural network, with a mean absolute error of 0.23 μm for surface roughness predictions. Their approach incorporated a classification step to identify and handle potentially problematic parameter combinations, improving the overall model accuracy. The system demonstrated better prediction performance than traditional simulation, particularly for complex parameters such as surface morphology, highlighting the potential of machine learning not just for prediction but also for optimizing process parameters while reducing the need for extensive empirical testing.

**Image-based Prediction**

Image-based prediction approaches represent a different approach for the determination of the resulting topography for laser surface texturing parameters, utilizing computer vision and deep learning to directly analyse surface morphology. Unlike traditional prediction methods that rely solely on processing parameters, these approaches utilize microscopy images of the processed surfaces to develop more comprehensive prediction models. By incorporating visual information about the resulting surface structures, these methods can capture subtle variations in texture quality and morphological features that might be missed by parameter-based approaches alone. Image-based machine learning techniques are particularly valuable for processes like LIPSS formation, where the quality assessment traditionally requires expert visual inspection of surface characteristics.

Wang et al. [36] demonstrated a hybrid machine learning approach that combines feature extraction through transfer learning with clustering and classification methods for optimizing femtosecond LIPSS. Their experiments utilized a Ti:sapphire laser amplifier system generating linearly polarized pulses of 35 fs duration and 800 nm wavelength at 1 kHz repetition rate, focused to a 30 μm spot size on Ti6Al4V specimens. The processing parameter space explored laser power (0.5-5 mW), scanning speed (0.5-1.5 mm/s), and number of scanning passes (1-5 times), generating 150 different sample conditions.

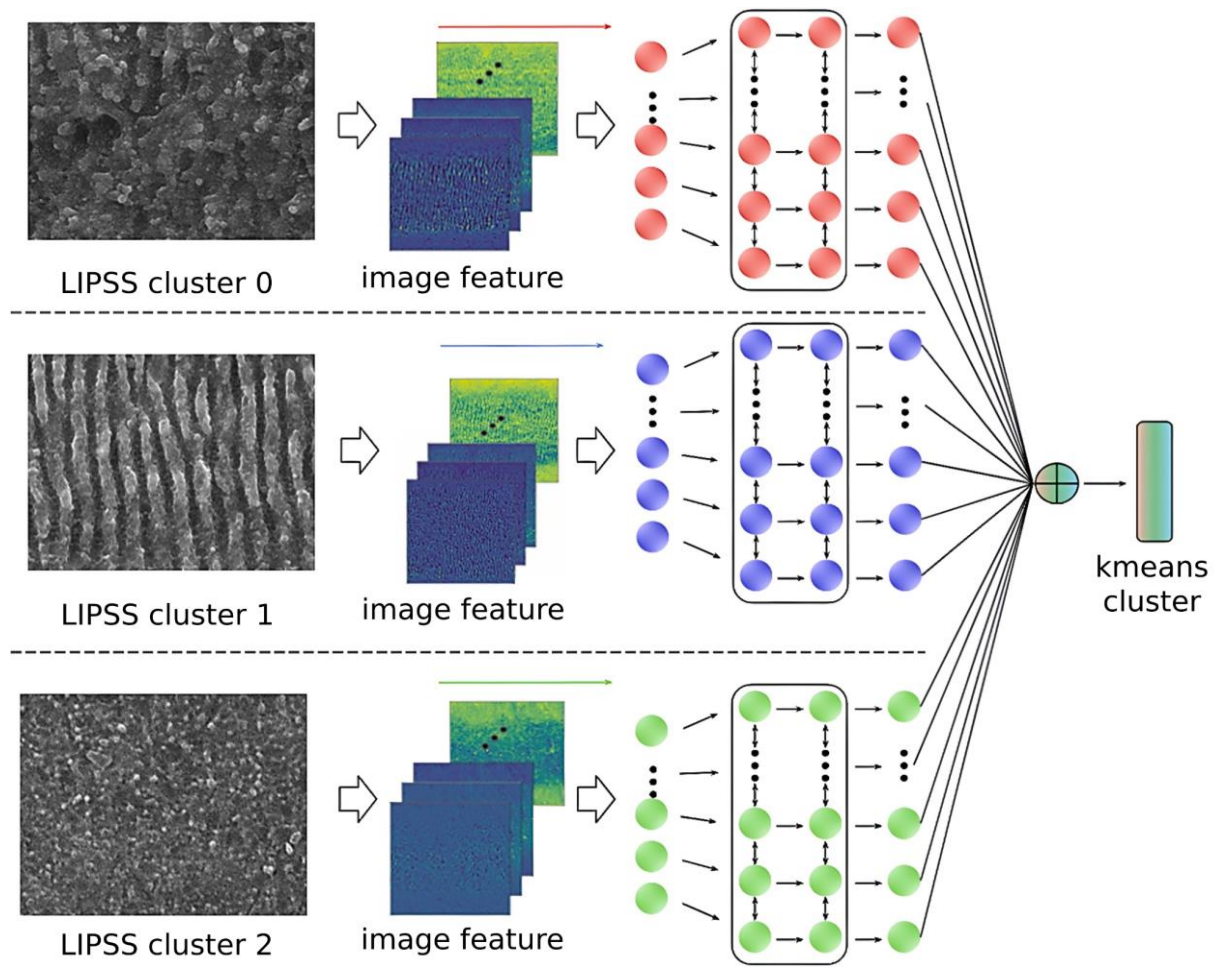

**Fig. 19.9:** Schematic depiction of workflow for the application of transfer learning to LIPSS quality assessment. SEM images of different LIPSS types are fed to pre-trained neural networks (VGG19) for feature extraction. Using K-Means clustering on the extracted features, the different types of LIPSS are automatically clustered into groups representing different LIPSS quality levels. This approach enhances clustering accuracy by replacing conventional dimensionality reduction with deep learning feature extraction. Reprinted from [36], Wang, B., Wang, P., Song, J., Lam, Y.C., Song, H. & Wang, Y. et al. (2022) A hybrid machine learning approach to determine the optimal processing window in femtosecond laser-induced periodic nanostructures. Journal of Materials Processing Technology, 308, 117716. Copyright 2022, with permission from Elsevier.

As shown in **Fig. 19.9**, their methodology first processes SEM images of different LIPSS morphologies showing three distinct structure types: disordered nanostructures (Cluster 0), well-defined periodic structures (Cluster 1), and fine scattered patterns (Cluster 2). For each SEM image, features are extracted through multiple processing layers using the VGG19 transfer learning model, represented by the stacked feature maps and subsequent neural network architectures. These extracted features were then fed into a k-means clustering algorithm, which automatically categorized the LIPSS patterns into the three morphology clusters. The neural network architecture for each cluster (shown in red, blue, and green) processed these features independently before combining them for final classification. The clustered data was used to train various classifiers, with Decision Trees achieving the highest accuracy of 96.7%, outperforming other methods like artificial neural networks (94.7%) and support vector machines (94.7%). This hybrid approach enabled automated determination of optimal processing windows for LIPSS fabrication, particularly identifying that good quality LIPSS could be achieved with laser powers between 2.5-3.5 mW and specific combinations of scanning times. By integrating multiple machine learning techniques (feature extraction, clustering, and classification), their method successfully mapped relationships between laser parameters and LIPSS quality, demonstrating how hybrid architectures can effectively address the complexities of laser surface texturing process optimization while reducing the need for time-consuming empirical parameter studies.

Li et al. demonstrated the combination of Response Surface Methodology (RSM) and ANN for predicting geometric features in femtosecond laser microfabrication on 4H-SiC wafers [37]. Their experimental setup employed a femtosecond laser (209 fs pulse duration, 1064 nm wavelength) focused through a 210 mm f-theta lens to create precise microgrooves (Fig. 19.10). The system utilized galvanometric scanning with spot sizes of 35 μm. Their study systematically varied three key processing parameters: laser power (2-8 W), scanning speed (200-1000 mm/s), and number of scanning passes (20-100).

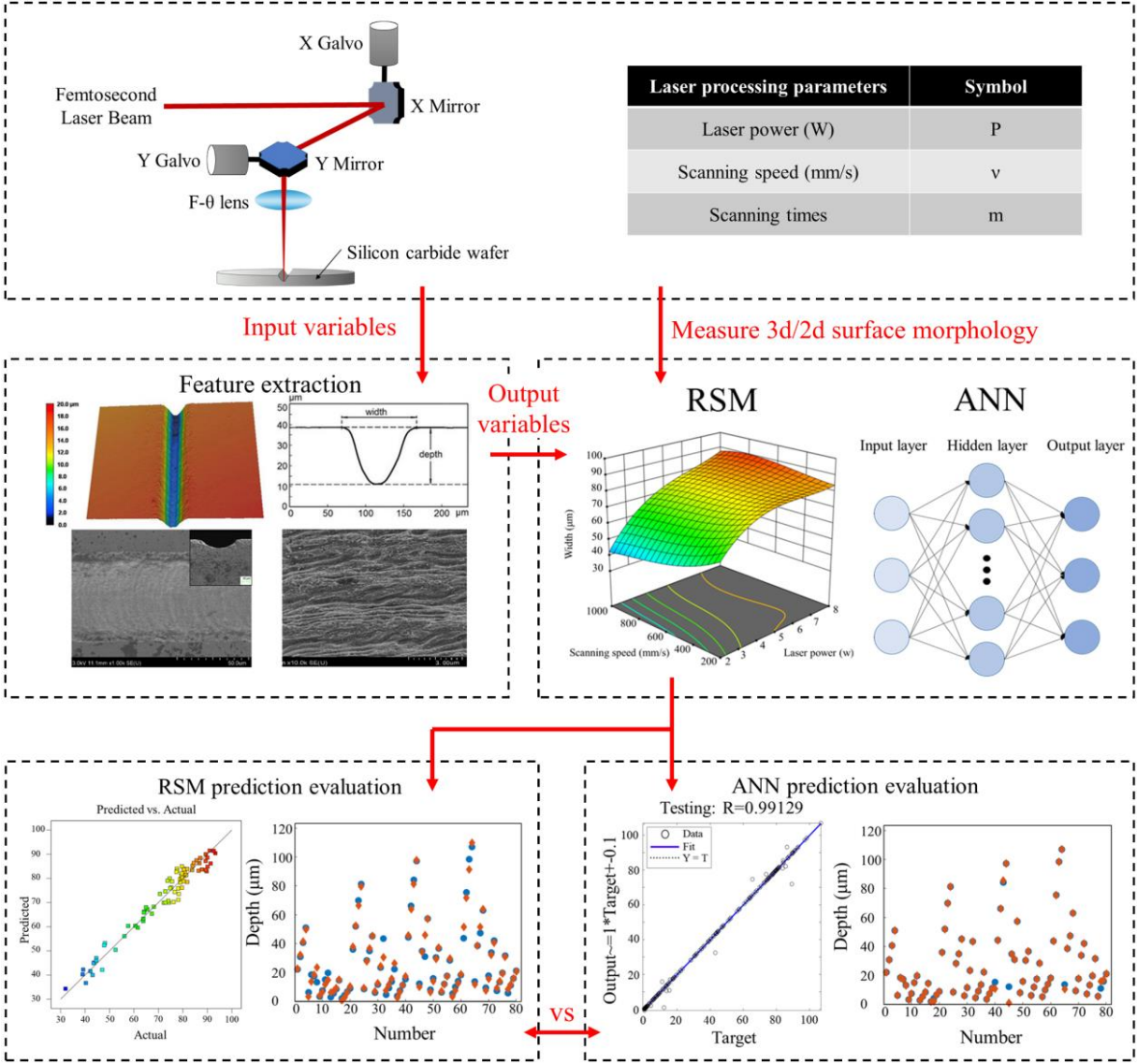

**Fig. 19.10**: Workflow showing RSM and ANN hybrid approach for predicting laser machining outcomes, combining feature extraction, modeling and evaluation stages. Adapted from [37], Li, X., Wang, H., Wang, B. & Guan, Y. (2022) Machine learning methods for prediction analyses of 4H–SiC microfabrication via femtosecond laser processing. Journal of Materials Research and Technology, 18, 2152–2165. Copyright 2022, with permission from Elsevier.

Their methodology demonstrated several key advantages of combining traditional statistical approaches with machine learning. The RSM analysis revealed that scanning speed had the strongest influence on depth, while laser power most significantly affected width. The parallel ANN implementation achieved superior prediction capabilities across all measured parameters, with an overall $R^2$ value of 0.9913 compared to RSM's varying performance ($R^2$ = 0.9815, 0.9592, and 0.4383 for depth, width, and surface roughness respectively). The ANN model particularly excelled at predicting surface roughness variations, where traditional RSM approaches struggled to capture the complex parameter interactions. This hybrid approach not only improved prediction accuracy but also reduced the number of experimental trials needed to optimize processing parameters.

## Feature Extraction for Multi-Sensor, Multi-Input Parameter Processing

In LST, predictive modeling can also require the integration of diverse sensor signals and process parameters to accurately estimate surface quality and geometrical outcomes. Multi-sensor setups typically capture time-series data from modalities such as infrared, acoustic, visual, and laser reflection signals, while multi-input parameters include variables like pulse energy, scanning velocity, number of exposures, and material properties.

To harness this complex and high-dimensional data, feature extraction plays a crucial role. Rolling statistical methods and specialized open-source libraries such as tsflex and tsfel are employed to convert raw time-series signals into compact, informative feature sets. These features capture both global trends and local variations in the process dynamics, enabling machine learning models to detect subtle correlations between sensor feedback and final surface characteristics.

Leyendecker et al. investigated this in the context of ultra-short pulse laser structuring to optimize process quality and efficiency [38]. A large-scale parameter study involving 60 experiments and 5,940 laser structures was conducted with five different preprocessing techniques (milling, grinding, polishing, die-EDM, and wire-EDM). During the experiments, multi-sensor process monitoring data (laser reflection, infrared, visual, and acoustic emissions) was collected at a high sampling rate of 100 kHz (Fig. 19.11). Their study highlighted the importance of preprocessing techniques and initial surface conditions on model accuracy. By extracting time-series features using rolling statistics, predictive performance was further improved, particularly for smoother surfaces.

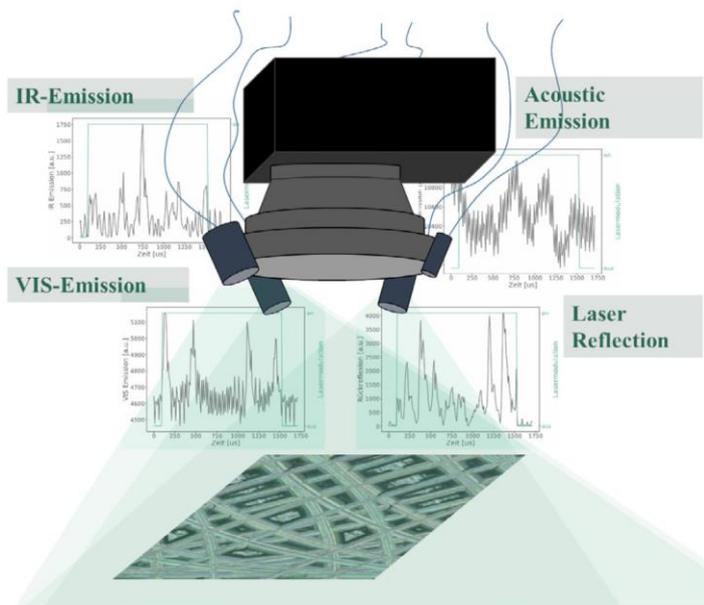

**Fig. 19.11:** Laser structuring process scheme with laser optic, process monitoring. Reprinted from [38], Leyendecker, L., Zuric, M., Nazar, M.A., Johannes, K. & Schmitt, R.H. (2023) Predictive Quality Modeling for Ultra-Short-Pulse Laser Structuring utilizing Machine Learning. Procedia CIRP, 117, 275–280. Copyright 2023 under Creative Commons BY 4.0 license. Retrieved from https://doi.org/10.1016/j.procir.2023.03.047.

Complementing this sensor-driven strategy, Liu et al. adopted a parameter-driven approach by developing a hybrid **GWO-BPNN** (Grey Wolf Optimization–Backpropagation Neural Network) model to predict femtosecond laser etching outcomes—specifically groove depth, width, and aspect ratio—based solely on laser process parameters [22].. The GWO algorithm was used to optimize network hyperparameters, such as the number of hidden neurons and training epochs, addressing the common challenges of slow convergence and local minima in traditional BPNNs. Their model achieved $R^2$ values exceeding 0.9 across all target variables, indicating high prediction accuracy and improved model stability.

As illustrated in Fig. 19.12, the GWO-BPNN model consistently outperformed the standalone BPNN. The predicted values (red) closely followed the experimental data (black), particularly for depth and width, while the BPNN predictions (blue) showed more pronounced deviations, especially for the aspect ratio. This visual comparison highlights the effectiveness of the GWO-enhanced model in capturing the underlying nonlinear relationships between process parameters and microstructural outcomes. The results confirm that metaheuristic optimization can significantly enhance machine learning-based prediction capabilities, even in the absence of sensor data, offering a fast and reliable solution for laser process modeling.

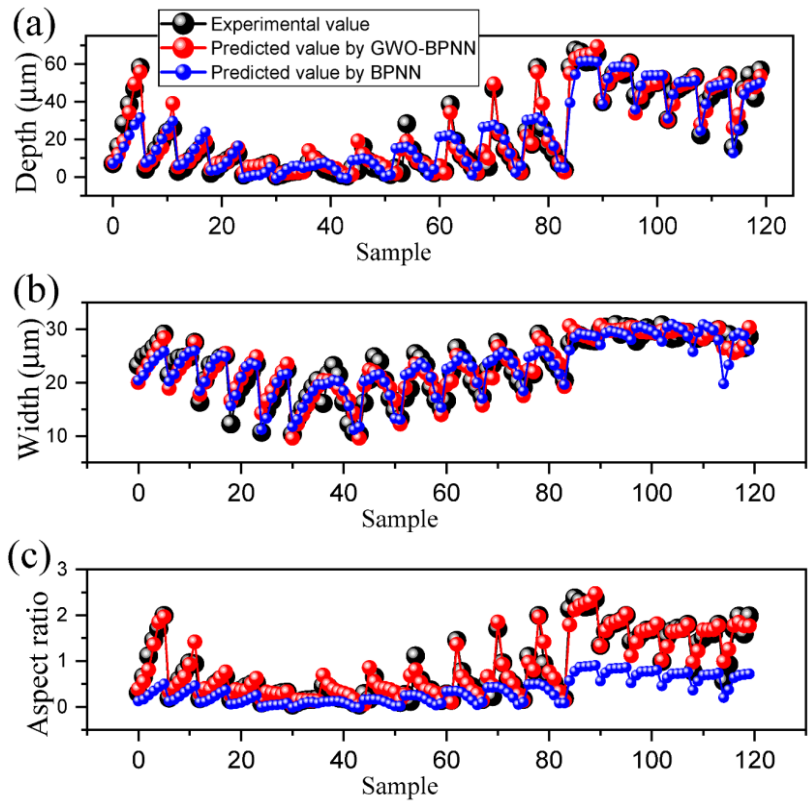

**Fig. 19.12:** Ilustration of the output variables obtained from GWO-BPNN and BPNN prediction and their comparison with the actual values. Compared with those of BPNN-predicted data, the curve behaviors of GWO-BPNN-predicted data were sufficiently consistent with the actual values, meaning that GWO was efficient in improving the prediction ability of the BPNN model. Adapted from [22], Liu, Y., Shangguan, D., Chen, L., Su, C. & Liu, J. (2024) Prediction of Femtosecond Laser Etching Parameters Based on a Backpropagation Neural Network with Grey Wolf Optimization Algorithm. Micromachines, 15(8). Copyright 2024 under Creative Commons BY 4.0 license. Retrieved from https://doi.org/10.3390/mi15080964.

## 2.3 Prediction of Surface Properties

Typical methods to predict surface properties, such as wetting behavior or heat flux, involve numerical models based on experimental data [39–42]. In such cases, finite element methods, often using software like COMSOL, are commonly employed to model the dynamics at the fluid-solid interface or heat exchange Vorgänge an de speziellen Geometrie. These models typically use microrelief measurement data or modeled surface reliefs as input. As a result, the accuracy of such simulations aligns well with experimental data only under limited boundary conditions. For example, in wetting predictions, these simulations primarily focus on topographical changes and do not account for variations in surface chemistry, which can significantly affect the surface's functional properties. [42]. Therefore, prediction based on machine learning methods can serve as an alternative, leading to more accurate predictions.

## Wetting Prediction

One approach to predict wetting behavior without a deep understanding of the interplay between surface chemistry and surface topography is demonstrated by Baronti et al., who used ANNs to predict the wetting behavior of surfaces treated with ultrafast lasers, specifically focusing on LIPSS [43]. The key AI algorithm employed is the Multi-Layer Perceptron (MLP), a type of feedforward neural network, which was trained to predict the wettability (measured by the static contact angle) of LIPSS-treated surfaces.

To build and train the ANN, the researchers used both experimental data and synthetic data generated through a Generative Adversarial Network (GAN). The GAN was employed as a data augmentation technique to significantly reduce the need for large sets of experimental data by generating realistic artificial LIPSS topographies (**Fig. 19.13.**a). These topographies, derived from Atomic Force Microscopy (AFM) scans, were converted into areal surface roughness parameters, which served as the input for the ANN model (**Fig. 19.13.**b).

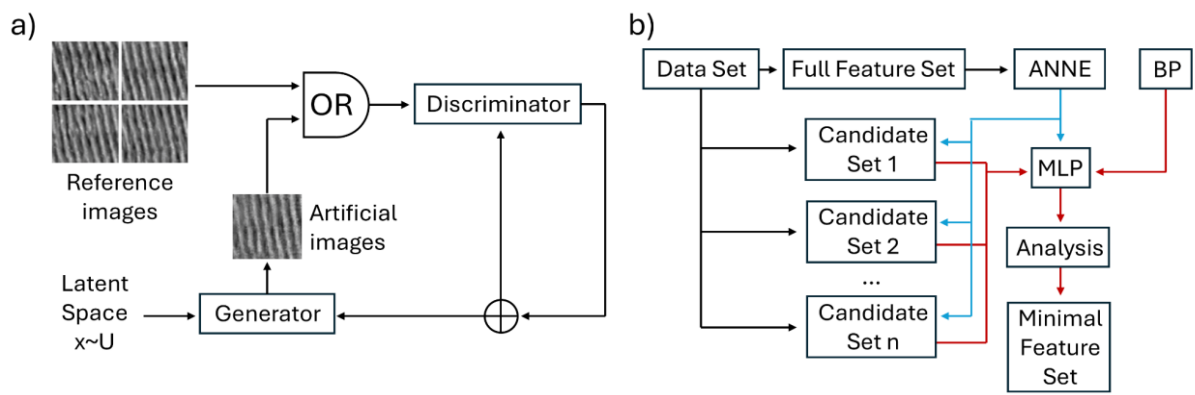

**Fig. 19.13:** a) Schematic representation of GAN used to generate artificial height maps/depth images based on experimental LIPSS topography data. b) Steps of feature relevance analysis split into two stages. In the first stage (blue lines), the ANNE procedure was used to optimise the MLP structure and generate candidate groups of surface parameters. In the second stage (red lines), the parameter groups were evaluated on the learning results of MLP (using BP training) and a final minimal group of relevant areal surface parameters is generated. Adapted from [43], Baronti, L., Michalek, A., Castellani, M., Penchev, P., See, T.L. & Dimov, S. (2022) Artificial neural network tools for predicting the functional response of ultrafast laser textured/structured surfaces. The International Journal of Advanced Manufacturing Technology, 119(5-6), 3501–3516. Copyright 2022 under Creative Commons BY 4.0 license. Retrieved from https://doi.org/10.1007/s00170-021-08589-9.

The primary task of the ANN was twofold: to classify laser processing disturbances (such as variations in Focal Offset Distance and Beam Incident Angle) and to predict the functional response of the treated surfaces, particularly their wettability. The ANN's prediction capability was evaluated through feature selection methods that reduced the number of input parameters to the most relevant ones, ensuring efficient and accurate performance. The feature selection process identified a reduced set of surface parameters, narrowing down the dataset from 21 to as few as 6 relevant parameters. The model achieved a high prediction accuracy, with the classification of laser processing disturbances reaching approximately 85%. In the wettability prediction task, the ANN provided a contact angle prediction with a root mean square error of 11 degrees, which is within the uncertainty of static water contact angle measurements (around 10 degrees).

Other works, such as those in [44], have used CNNs as deep learning models to predict the contact angle based on topography and laser parameters, or to track the development of contact angles over time [45]. In the approach shown in **Fig. 19.14**, a CNN model was applied to predict the contact angle of a surface patterned with rectangular or cylindrical pillars. The model uses a three-dimensional input descriptor, represented as a unit cell with surface topography information, to predict the CA of a droplet placed on the surface. The CNN consists of five convolutional blocks followed by fully connected layers, which process the surface topography data and generate the predicted contact angle (θ_CNN). The network is trained to match the experimental contact angle (θ_exp), with a demonstrated high correlation ($R^2 = 0.88$) between the predicted and experimental values.

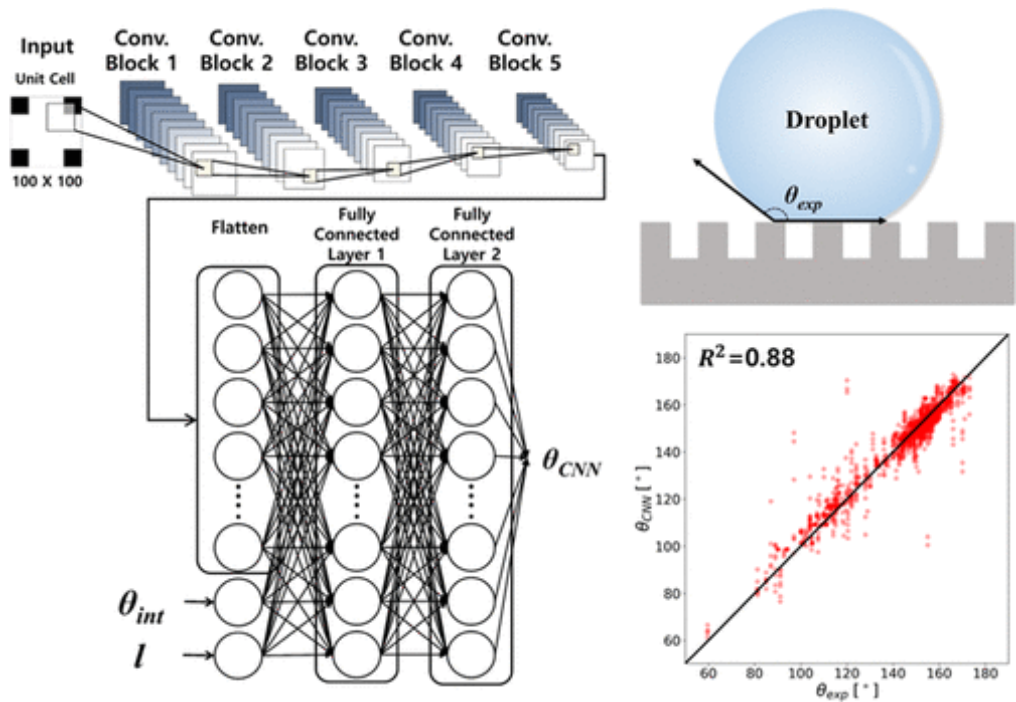

**Fig. 19.14:** The figure illustrates a convolutional neural network (CNN) model used to predict the contact angle (CA) of a surface patterned with rectangular or cylindrical pillars. The input consists of a 100x100 unit cell representing the surface topography. The network includes five convolutional blocks followed by fully connected layers. The CNN model outputs the predicted contact angle (θ_CNN), which is compared to the experimental contact angle (θ_exp). The plot on the bottom right shows a strong correlation ($R^2 = 0.88$) between the predicted and experimental values, demonstrating the model's effectiveness in predicting wettability based on surface features. Reprinted with permission from [44], Choi, S., Kim, K., Byun, K. & Jang, J. (2023) Machine-Learning Approach in Prediction of the Wettability of a Surface Textured with Microscale Pillars. Langmuir, 39(48), 17471–17479. Copyright 2023 American Chemical Society.

Laser-textured surfaces typically undergo a gradual transition from hydrophilic to hydrophobic states, limiting their practical applications. To address this, Zhang et al. developed a multimodal deep learning framework using a CNN, which combines multiple data types, such as SEM and EDX images, with temporal data, to predict how surface wettability evolves over time for various micro/nanostructures. The framework first overlays the SEM and EDX images with time-related information to create a multimodal input dataset (1st step in **Fig. 19. 15**). This data is then processed by a deep learning model consisting of several convolutional and max-pooling layers to extract relevant features (2nd step in **Fig. 19. 15**). The model outputs a contact angle curve, predicting wettability changes over time (3rd step in **Fig. 19. 15**). Visualization of the convolution layers confirmed that the model was identifying the necessary features, and the framework successfully optimized surfaces to achieve faster evolution times and larger contact angles. The output graphs show the actual versus predicted contact angle curves for both the training and validation sets, with a high degree of accuracy.

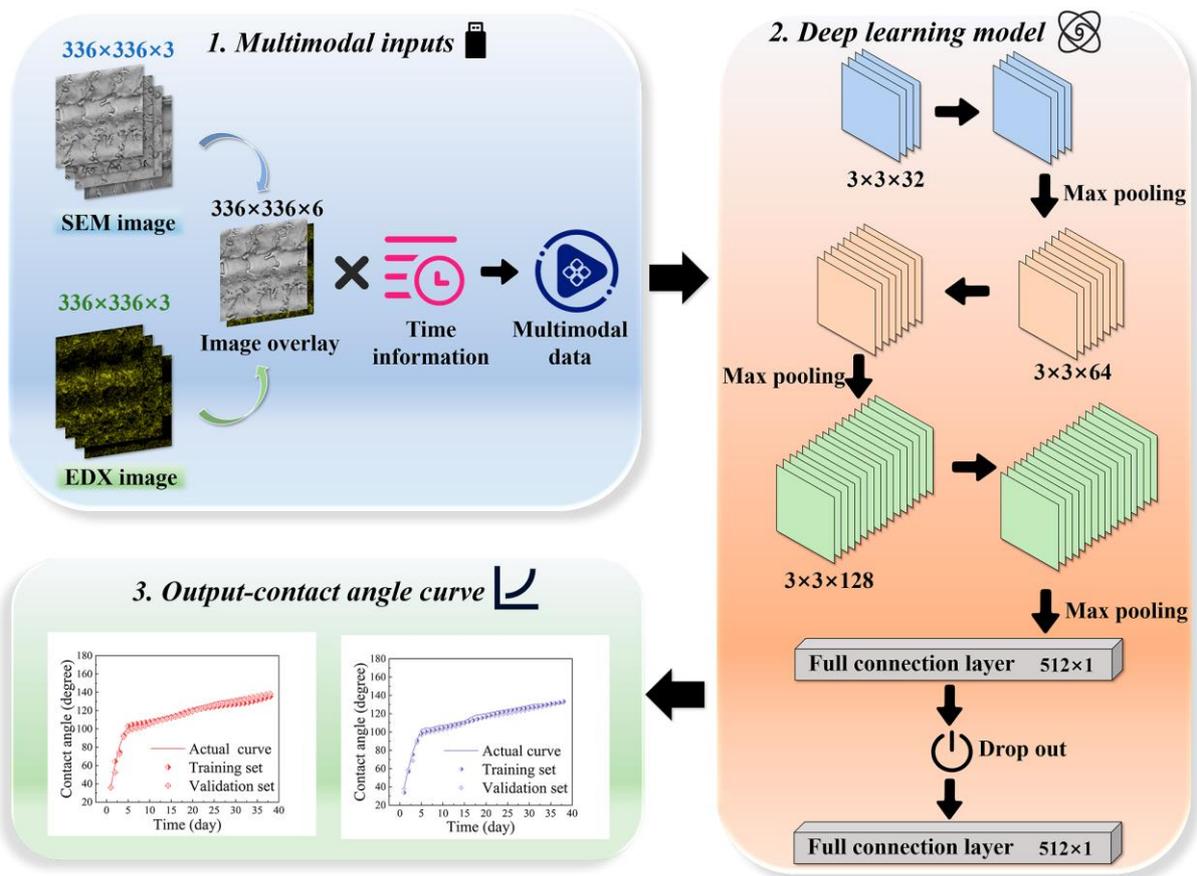

**Fig. 19. 15:** Display of a multimodal deep learning framework for predicting the evolution of surface wettability. 1.) SEM and EDX images are combined with time-related data, creating a multimodal input. 2.) This data is processed through a CNN model with several convolutional and max-pooling layers, followed by fully connected layers. 3.) The output is a contact angle curve that predicts the wettability change over time. The graphs display actual versus predicted contact angles for both the training and validation sets, demonstrating the model's accuracy in forecasting wettability evolution. Reprinted with permission from [45], Zhang, Z., Yang, Z., Zhao, Z., Liu, Y., Wang, C. & Xu, W. (2023) Multimodal Deep-Learning Framework for Accurate Prediction of Wettability Evolution of Laser-Textured Surfaces. ACS Applied Materials & Interfaces, 15(7), 10261–10272. Copyright 2023 American Chemical Society.

### Prediction of Reflection Coefficient and Antibacterial Properties

Predicting surface properties using machine learning is not only important for wetting prediction but also essential for a wide range of applications. In a comprehensive study, Na et al. [45] demonstrated that a GAN could be used to generate artifical SEM images of LIPSS under unknown parameter combinations. The model takes laser fluence and scanning speed as input values, and based on these generated images, it predicts the reflective spectra [46]. In **Fig. 19. 16**, the model they used is shown and works as follows.

Part A (Topology to Scalar Reflectance): The input is a surface topography image (C, H, W), which has dimensions (1, 256, 256). This image is processed through a ResNet 152 model, which uses a modified Resblock architecture with layers consisting of convolutions, batch normalization, and Leaky ReLU activations. The model outputs a 2048-dimensional vector, which is then used to predict the scalar reflectance at 550 nm. The overall architecture is designed to extract features from the surface image and output a single scalar value representing reflectance.

Part B (Topography to Reflectance Profile): This part follows a similar process, but the model is designed to predict a reflectance profile instead of a single scalar value. Again, the input is a surface topography image processed by the ResNet 152. The output of the ResNet is a 2048-dimensional vector. This vector is then passed through several LSTM (Long Short-Term Memory) cells, which handle sequential data and are responsible for generating a reflectance profile with 41 dimensions. Each LSTM layer has fixed weights and outputs vectors that are processed further by the LSTM network. Additionally, dropout is used in the architecture for regularization, helping prevent overfitting.

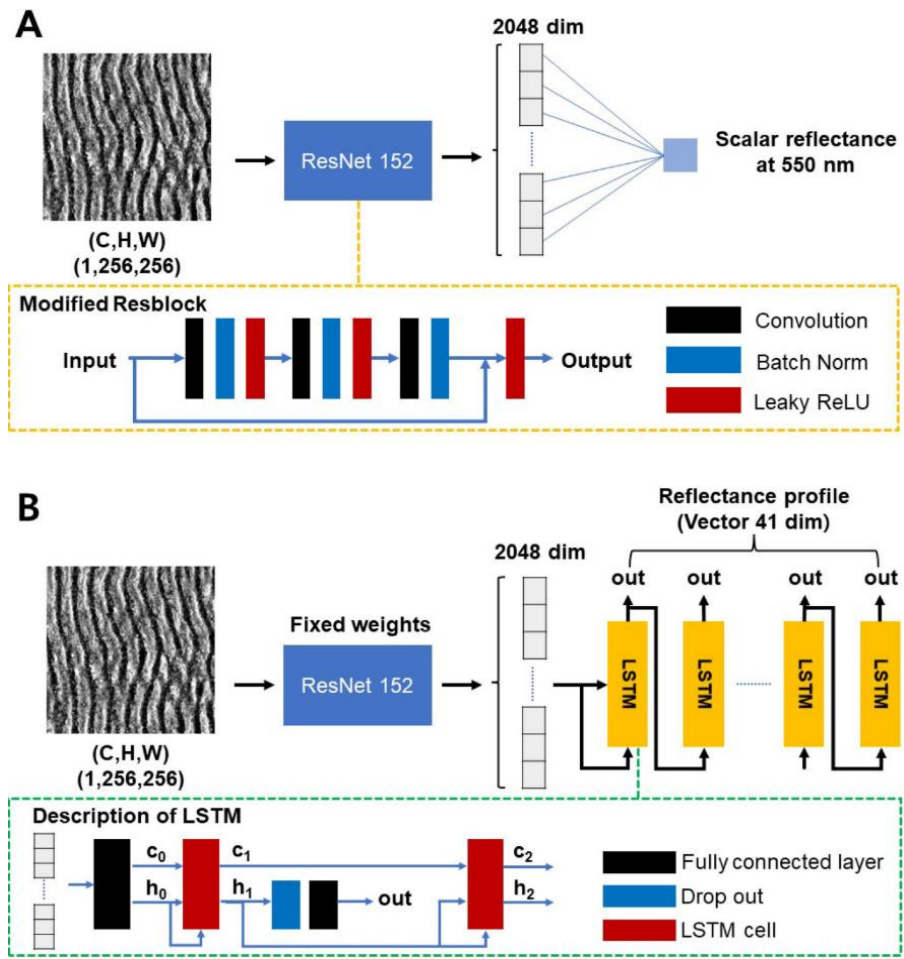

**Fig. 19. 16**: Overall structures of the reflectance prediction models. Model A predicts scalar reflectance at 550 nm. Model B predicts 41-dimension reflectance spectrum from 350 nm to 750 nm. Reprinted from [46], Na, H., Yoo, J. & Ki, H. (2022) Prediction of surface morphology and reflection spectrum of laser-induced periodic surface structures using deep learning. Journal of Manufacturing Processes, 84, 1274–1283. Copyright 2022, with permission from Elsevier.

In summary, the model utilizes both a deep convolutional network (ResNet 152) to extract features from surface images and LSTM cells to predict a time-dependent or sequential reflectance profile for predicting surface reflectance properties based on topographical data. The Pattern period and ripple width were predicted with 98.2 and 94.6% accurarcy, respectively and the reflection spectra of the surfaces were predicted with less than 4% norm error.

A simpler way to predict the reflection coefficient of a femtosecond laser-processed surface in real-time, replacing traditional spectrometer measurements, was demonstrated by Liu et al. [47]. The method uses a Support Vector Regression (SVR) model, which is trained with laser processing parameters and image features extracted from SEM images to predict the laser ablation diameter. The trained SVR model then predicts the ablation diameter from new SEM images, and the reflection coefficient is computed accordingly. This approach achieves a reflection coefficient estimation accuracy of over 90% and was mainly developed to measure blackened shells of X-ray imaging sensors.

Another interesting study by Zhang et al. explores the forecast of antibacterial properties described **Fig. 19. 17** applying femtosecond laser treatment on PEEK [48]. An antibacterial dataset was created by collecting SEM images and corresponding antibacterial rates from published literature. The dataset included 63 SEM images from various materials with surface structures that affect antibacterial properties against S. aureus and E. coli. The antibacterial rate was defined as the percentage decrease in bacterial adhesion compared to smooth surfaces. This dataset focused on physical surface structures, excluding chemical composition and wettability factors. A Random Forest Classifier (RFC) machine learning model were trained on this dataset to predict the antibacterial enhancement due to surface structure. Finally, PEEK surfaces were processed using the femtosecond laser technique shown in Step 1, and their antibacterial rates were predicted using the RFC model. The best-predicted samples were selected for in vitro antibacterial testing against E. coli and S. aureus. The results validated both the antibacterial effects of the surfaces and the accuracy of the RFC model's predictions with an overall prediction accuracy exceeding 90%.

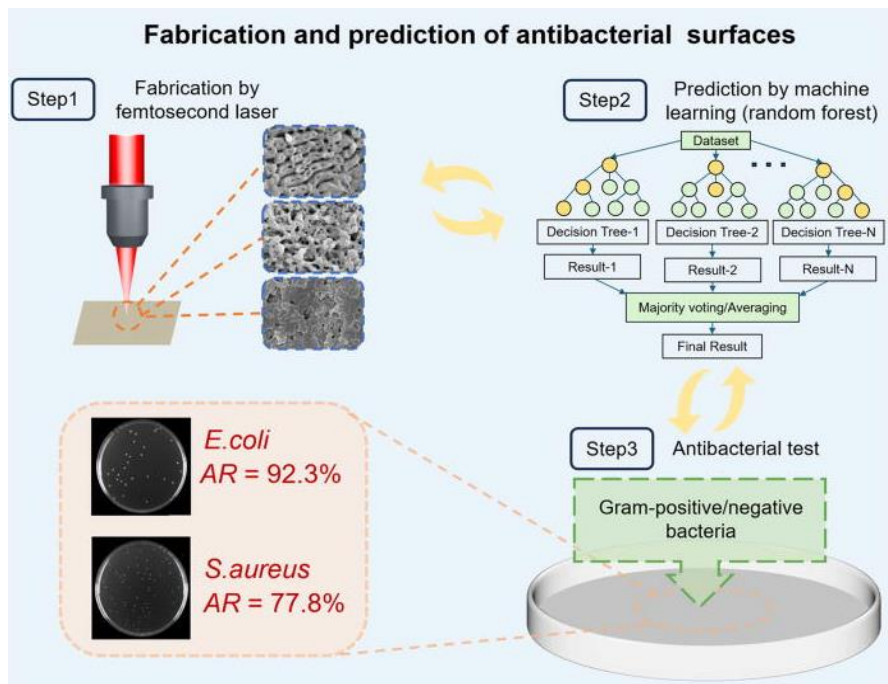

**Fig. 19. 17:** Fabrication of antibacterial surfaces through femtosecond laser treatment, and prediction of antibacterial rate by machine learning, applying random forest algorithm. Reprinted from [48], Zhang, W., Gao, K., Li, X., Li, H., Liao, J. & Xuan, S. et al. (2024) Enhancing antibacterial properties of PEEK surfaces: Laser-induced and machine-learning assessed. Applied Physics Letters, 125(15), with the permission of AIP Publishing.

# 3 Process Control

In this section, monitoring and process control techniques that leverage machine learning and artificial intelligence are discussed. The results are categorized into two main approaches: post-processing control and emission-based control.

## 3.1 Post-process Control

Post-processing monitoring is a crucial step in ensuring high-quality outcomes, especially for applications demanding micron-level accuracy. This approach typically involves analyzing completed parts to assess surface finish, a critical factor in various fields such as tribology, medical device manufacturing, and glass processing [49–51]. Common methods for post-process monitoring of laser-textured surfaces include imaging techniques like white light microscopy and scanning electron microscopy (SEM), as well as topography measurement techniques such as white light interferometry and confocal microscopy [52–55]. While there are widely used tools to analyze the surface features such as Gwyddion [56] or MountainsMap® [57], these software packages often suffer from limited automation capabilities. This represents a challenge for the analysis of large datasets and the integration of the topographical evaluation into a closed-loop system for real-time process control. Recently, the open-source Python library *surfalize* was proposed to address these gaps [58]. The library offers algorithms for the most common post-processing operations, such as filtering, detrending and denoising, as well as a large set of roughness parameters from ISO 25178. Additionally, it enables fast batch-processing of large datasets through a dedicated batch interface, leveraging multi-core processing. Through its full integration into the Python ecosystem, it allows for the direct interconnection with typical ML frameworks as well as for custom evaluation algorithms on the topographic data, integrating directly into the ML pipeline.

Measurement devices such as SEMs cannot be integrated into laser systems due to their size and complexity, and topography measurement systems are also not easily integrable or user-friendly in this context. Examples exist in the literature of topography measurement systems that integrate a probe beam directly into the machining laser's beam path [14,59]. However, since these systems are not applied to laser surface texturing or do not involve the use of machine learning algorithms, these publications will not be discussed further in this chapter.

Cameras equipped with appropriate magnifications, integrated directly into the beam path of the laser system, present a practical solution for in-machine post-monitoring. This integration significantly simplifies automation. However, in the past, the limited informational depth of microscopic images compared to topography measurement systems posed challenges. Recent approaches leverage microscopic images of laser-textured surfaces captured by cameras—either integrated into the machine or standalone—and use machine learning algorithms for process monitoring. These algorithms not only analyze the data but can also be incorporated into a closed-loop system, enabling real-time feedback and process control. Some of these systems have already been successfully implemented, showcasing their potential to enhance the automation and efficiency of post-process monitoring.

**Pattern Recognition for (Sub-)Micrometer Processing**

A key advancement in post-processing monitoring is the use of ML to classify self-organized surface structures produced during ultrashort-pulse laser processing. These structures, including LIPSS, craters, and microscale depressions, are highly dependent on complex interactions between laser parameters and material properties. In the study of Thomas et al. [60], a novel ML-based classification method was developed using light microscopy images. Four distinct surface types—polished reference, LIPSS, CRATER, and MICRO—were fabricated on hot work tool steel (HWTS) and stainless steel (SS) substrates using a 300-fs laser system. Digital light microscopy images were captured under optimized magnifications, creating a dataset of 1,000 HWTS images for training (**Figure 2.18**). An open-source tool, Teachable Machine, was employed to develop and train the ML algorithm, which achieved 100% accuracy in classifying validation images. The algorithm demonstrated robustness when applied to test datasets from both HWTS and SS substrates, as well as to images captured at varying optical magnifications. A web-based application was developed for real-time classification, showcasing the system's potential for integration into industrial workflows. By automating surface structure classification, this approach reduces reliance on costly, time-intensive methods such as SEM and provides a scalable solution for inline quality monitoring.

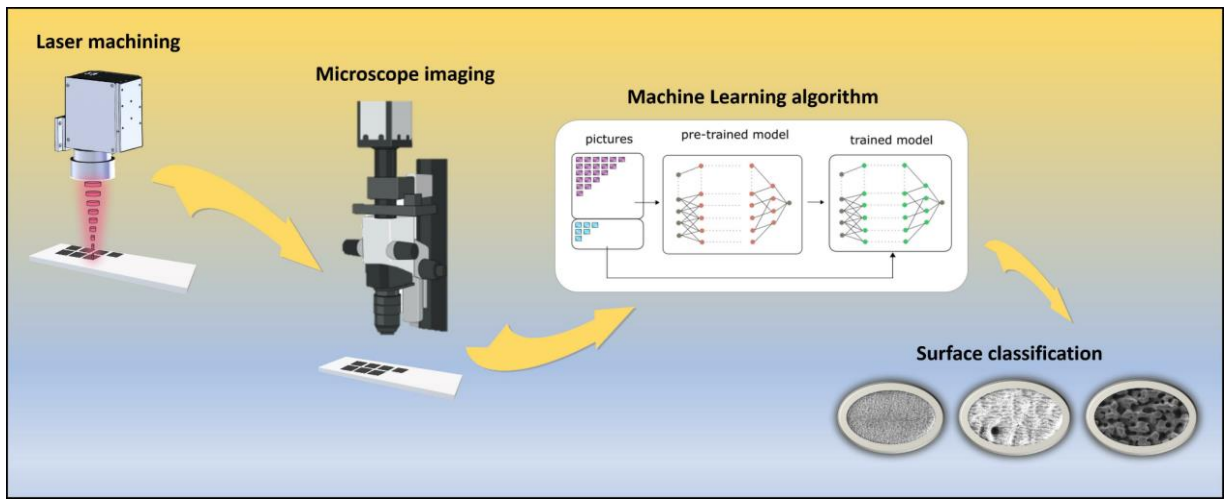

**Fig. 19. 18**: Graphical abstract from the work of Thomas et. al, indicating the procedure for classifying of LIPSS structures. Reprinted from [60], Thomas, R., Westphal, E., Schnell, G. & Seitz, H. (2024) Machine Learning Classification of Self-Organized Surface Structures in Ultrashort-Pulse Laser Processing Based on Light Microscopic Images. Micromachines, 15(4). Copyright 2024 under Creative Commons BY 4.0 license. Retrieved from https://doi.org/10.3390/mi15040491.

In another work by Mills et al. [61] the application of CNNs to identify and predict key parameters of the laser machining process directly from images of the machined surfaces was studied. The CNN approach bypasses the need for a detailed understanding of the photon–atom interactions, focusing instead on pattern recognition to infer parameters such as material type, laser fluence, and the number of laser pulses. By enabling real-time analysis, this method paves the way for precision feedback loops capable of correcting deviations during the machining process.

The experimental setup they used incorporates a femtosecond Ti:sapphire laser system generating 150-fs pulses at 800 nm, with fluence and pulse count precisely controlled. A camera captures images of the sample surface during machining, which are processed and analyzed by the CNN. Training the CNN required collecting a comprehensive dataset of 1,800 images across silica and nickel substrates, each corresponding to distinct combinations of laser parameters. Images were preprocessed to enhance signal-to-noise ratios and downscaled to 28x28 pixels for efficient analysis. The resulting false-color images and accuracy of the model are shown in **Fig. 19.19**.

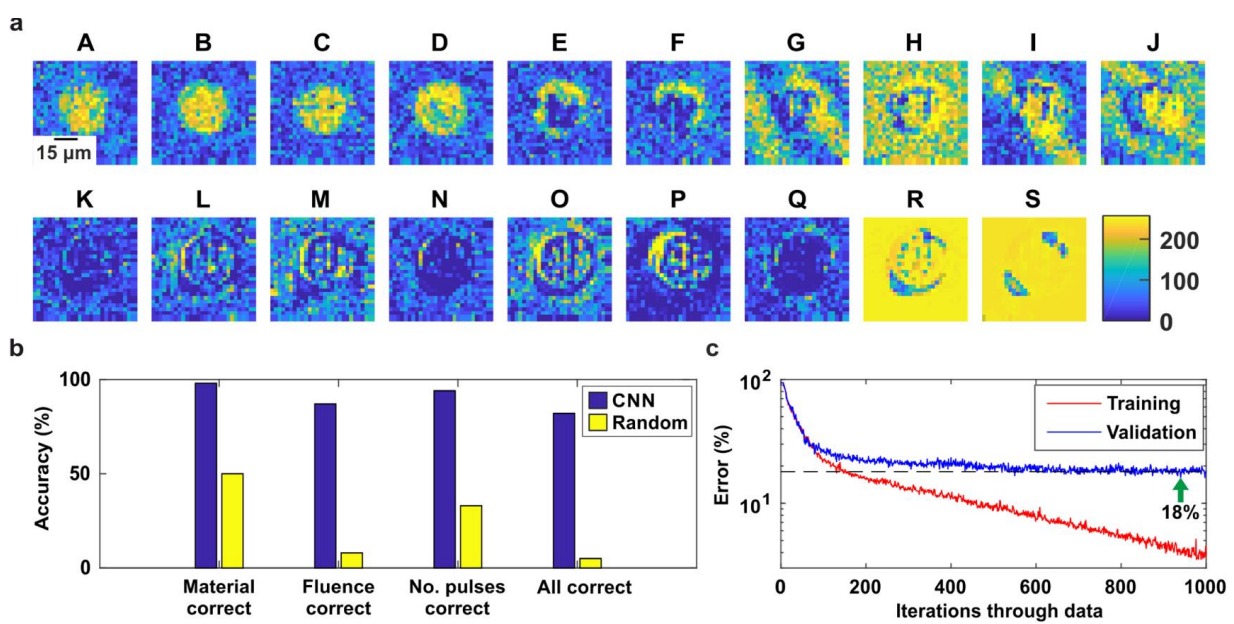

**Fig. 19.19:** (a) Examples of experimental pixel-binned images from the validation dataset, for each category, where the CNN correctly predicted the category. The detected intensity is represented in false-color from 0 to

255. (b) Prediction accuracy of the CNN, as compared with the accuracy of a random number generator. (c) Fitting error for the CNN for training and validation (minimum at 18%) datasets during training. Reprinted from [61], Mills, B., Heath, D.J., Grant-Jacob, J.A., Xie, Y. & Eason, R.W. (2019) Image-based monitoring of femtosecond laser machining via a neural network. Journal of Physics: Photonics, 1(1), 15008. Copyright 2018 under Creative Commons BY 3.0 license. Retrieved from https://doi.org/10.1088/2515-7647/aad5a0.

The CNN architecture was designed with three convolutional layers followed by a fully connected layer, optimized through iterative training over 1,000 epochs. Once trained, the CNN achieved an 82% accuracy in predicting all experimental parameters within milliseconds. The network demonstrated the ability to distinguish subtle variations in surface features caused by changes in fluence or pulse count, even when experimental noise introduced artifacts like debris or kerf formation. Validation tests showed high predictive accuracy, with clear identification of parameter categories despite significant overlap in visual features. The system's ability to process images in under 10 milliseconds makes it viable for real-time monitoring and corrective actions.

### Monitoring of Beam Alignment and Layer Ablation

AI-enhanced light microscopy can monitor not only surface topography but also laser beam behavior by visually analyzing the workpiece during machining. In the work by Yunhui Xie et al. [62] CNNs are trained to detect unintended beam translations and rotations and provide feedback capable of halting machining upon reaching specified completion points. This is exemplified in **Fig. 19.20**a, which schematically illustrates the experimental setup for real-time feedback.

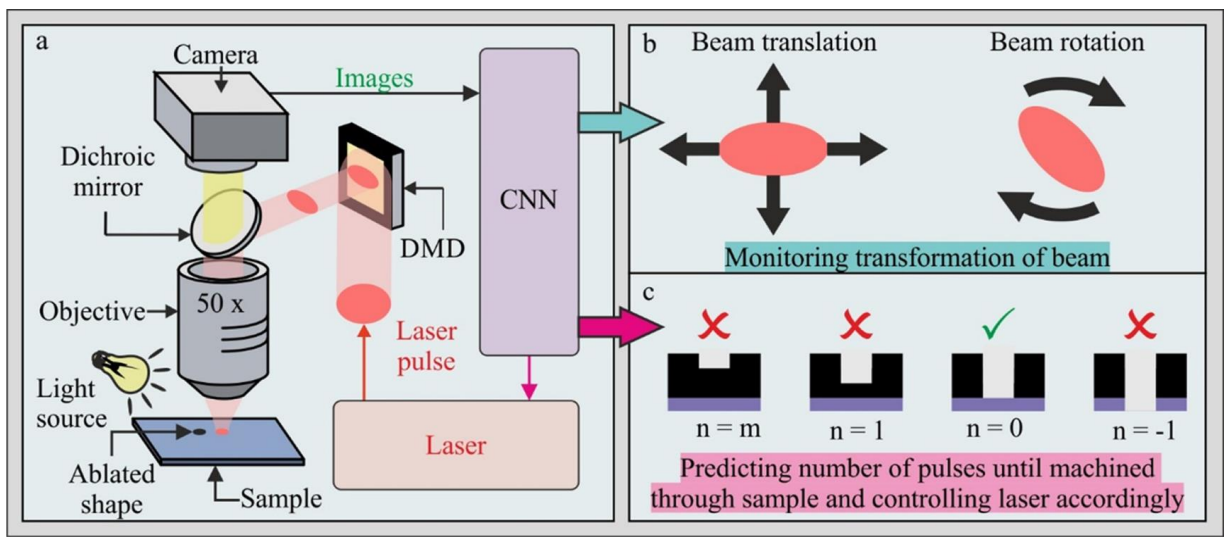

**Fig. 19.20**: Schematic of (a) experimental setup for real-time closed-loop feedback, and the concepts of (b) detecting a beam transformation (translation and rotation) and (c) predicting the remaining number of pulses until breakthrough of a thin film. Reprinted from [62], Xie, Y., Heath, D.J., Grant-Jacob, J.A., Mackay, B.S., McDonnell, M.D.T. & Praeger, M. et al. (2019) Deep learning for the monitoring and process control of femtosecond laser machining. Journal of Physics: Photonics, 1(3), 35002. Copyright 2019 under Creative Commons BY 3.0 license. Retrieved from https://doi.org/10.1088/2515-7647/ab281a.

The study shows how CNNs detect beam translations as small as $91 \pm 15$ nm, far exceeding the resolution of the camera used. **Fig. 19.20**b highlights the accuracy of detecting such sub-pixel displacements by simulating random beam transformations with a digital micromirror device (DMD). The neural networks were also successfully applied to detect simultaneous translations and rotations of the laser beam. Further experiments demonstrated real-time control of machining through thin copper films (~450 nm). The CNN accurately predicted the remaining pulses required for complete material penetration, halting the laser precisely at the breakthrough point. The feedback system, shows a significant reduction in error compared to naïve mean predictions, with CNN-based predictions outperforming baseline models by ~23%.

## 3.2 Emission-based Process Control

The interaction of laser and material during ultrashort pulse laser ablation involves multiple stages, including ablation, vaporization, plasma formation, and redeposition, depending on the applied energy density. For a detailed explanation of laser-material interactions on USP timescales, readers are directed to references [63–65]. **Fig. 19.21** illustrates an example of process signals emitted from the micromachining zone. Detecting phenomena such as acoustic shock waves, plasma plumes, and laser reflections during processing requires the integration of various sensors, including acoustic emission sensors, photodiodes, spectrometers, pyrometers, and thermal cameras, tailored to the specific dynamics of the micromachining process [66,67].

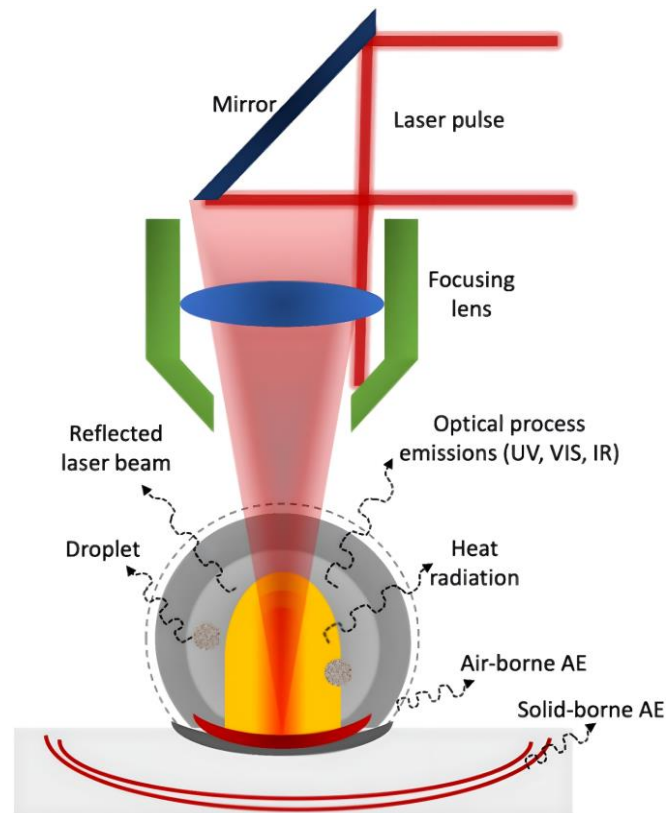

**Fig. 19.21:** Schematic representation of the process signals emitted during laser micromanufacturing. Due to their origin directly from the interaction zone they form the starting point for process and quality monitoring. Adapted from [14], Yildirim, K., Nagarajan, B., Tjahjowidodo, T. & Castagne, S. (2025) Review of in-situ process monitoring for ultra-short pulse laser micromanufacturing. Journal of Manufacturing Processes, 133, 1126–1159, Copyright 2024, with permission from Elsevier.

Machine learning approaches for analyzing emissions during laser surface texturing to control the laser process represent a highly novel and underexplored area of research. Notably, studies by Grant-Jacob et al. have primarily focused on imaging-based methods for plasma analysis [68]. This focus is particularly interesting, given the widespread use of acoustic emissions and spectral analysis in related fields, where the recognition of acoustic or spectral patterns is often considered relatively straightforward. However, most existing studies rely predominantly on amplitude values of the analyzed signals for process control, suggesting significant potential for more advanced, machine-learning-driven methodologies [69,70]. One possible limitation could be the increased computational demands and longer processing times required by machine learning algorithms compared to conventional feedback loops, which may impact the practicality of real-time process control.

**Process control by plasma imaging**

A novel method for real-time imaging of the surface of a silicon sample during femtosecond laser machining was introduced by Grant-Jacob et al. [71] .They applied convolutional generative adversarial networks (cGANs) to predict the surface morphology before and after a laser pulse, using only images of the plasma generated during machining. It was found that this approach overcame the challenge of plasma opacity, providing indirect but accurate visualizations of the surface.

To achieve this, a femtosecond laser was used to machine a silicon sample, and plasma images were captured perpendicular to the laser axis. The researchers trained two neural networks: one to predict the surface before the laser pulse and another to predict the surface after machining. They collected over 4,000 paired plasma and surface images for training and validated the models using independent datasets. It was observed that the trained networks achieved high accuracy in predicting the surface morphology.

During real-time implementation, the models were integrated into an automated system. The networks processed plasma images and generated predictions of the surface within approximately one second. It was found that the plasma's spatial structure strongly correlated with the surface morphology before machining. For post-machining surfaces, the neural network appeared to simulate the effects of the laser pulse based on the plasma data, enabling accurate predictions of the machined surface. The experimental setup for plasma imaging is shown in **Fig. 19.22**

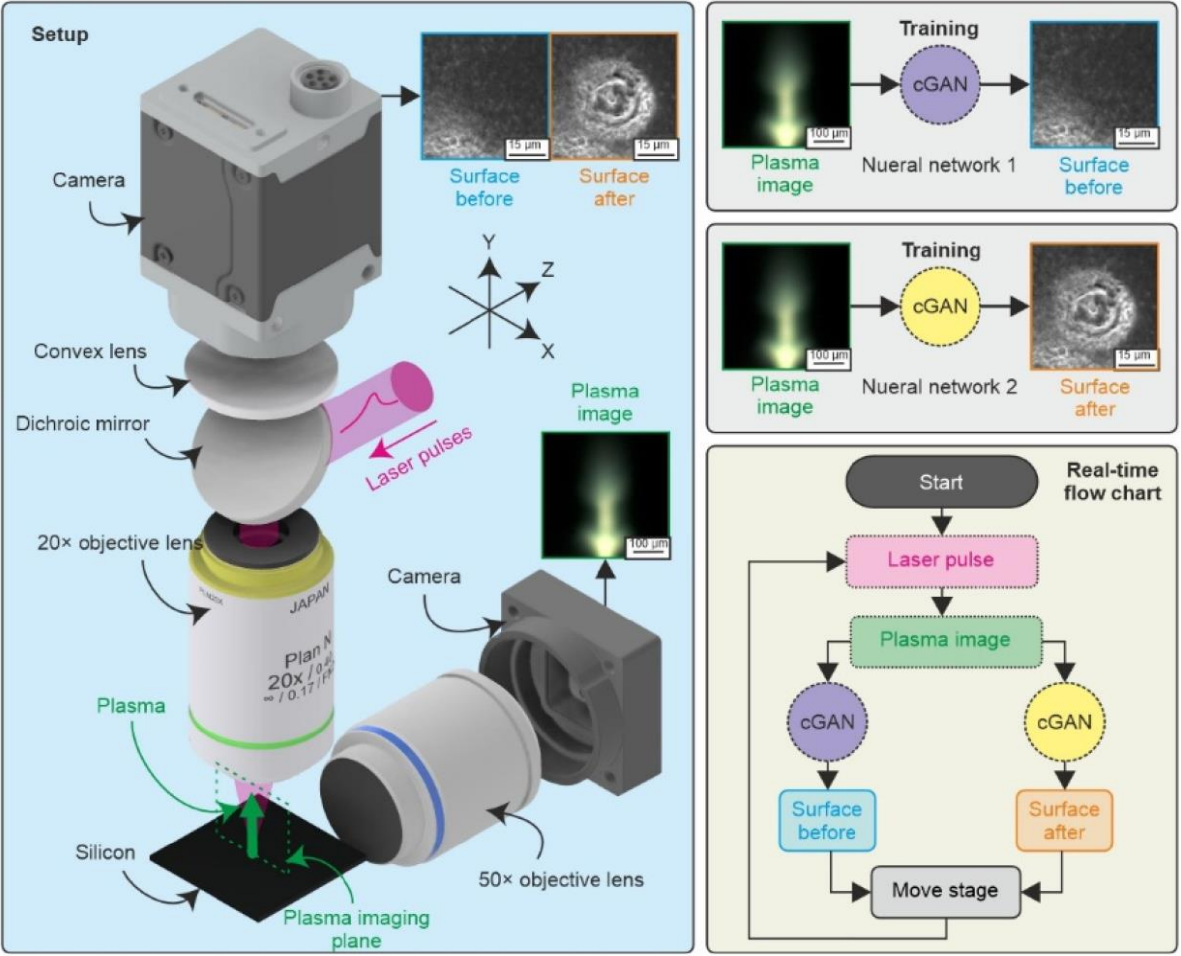

**Fig. 19.22**: Schematic of the experimental setup along with an example set of experimental plasma images and associated experimental images of the laser machined sample before and after the laser pulse. For this work, the two neural networks were run in real-time, hence providing a live image of the sample during machining. Reprinted from [71], Grant-Jacob, J.A., Mills, B. & Zervas, M.N. (2023) Live imaging of laser machining via plasma deep learning. Optics Express, 31(25), 42581–42594. Copyright 2023 under Creative Commons BY 4.0 license. Retrieved from https://doi.org/10.1364/OE.507708.

In a subsequent study, the setup was employed to automatically prevent the laser beam from scanning beyond control [72]. CNNs trained on images of emitted plasma were utilized to identify material boundaries in real time and to inhibit the laser beam from scanning beyond those boundaries. Data captured by two cameras during laser-material interactions were used: one camera recorded the surface, while the other captured plasma emissions (**Fig. 19.23**). The plasma images were processed through a CNN, which classified different regions—such as material, boundary, and air—based on distinct plasma characteristics. Boundary detection was achieved with a latency of approximately 100 milliseconds, enabling dynamic control of the laser scanning process.

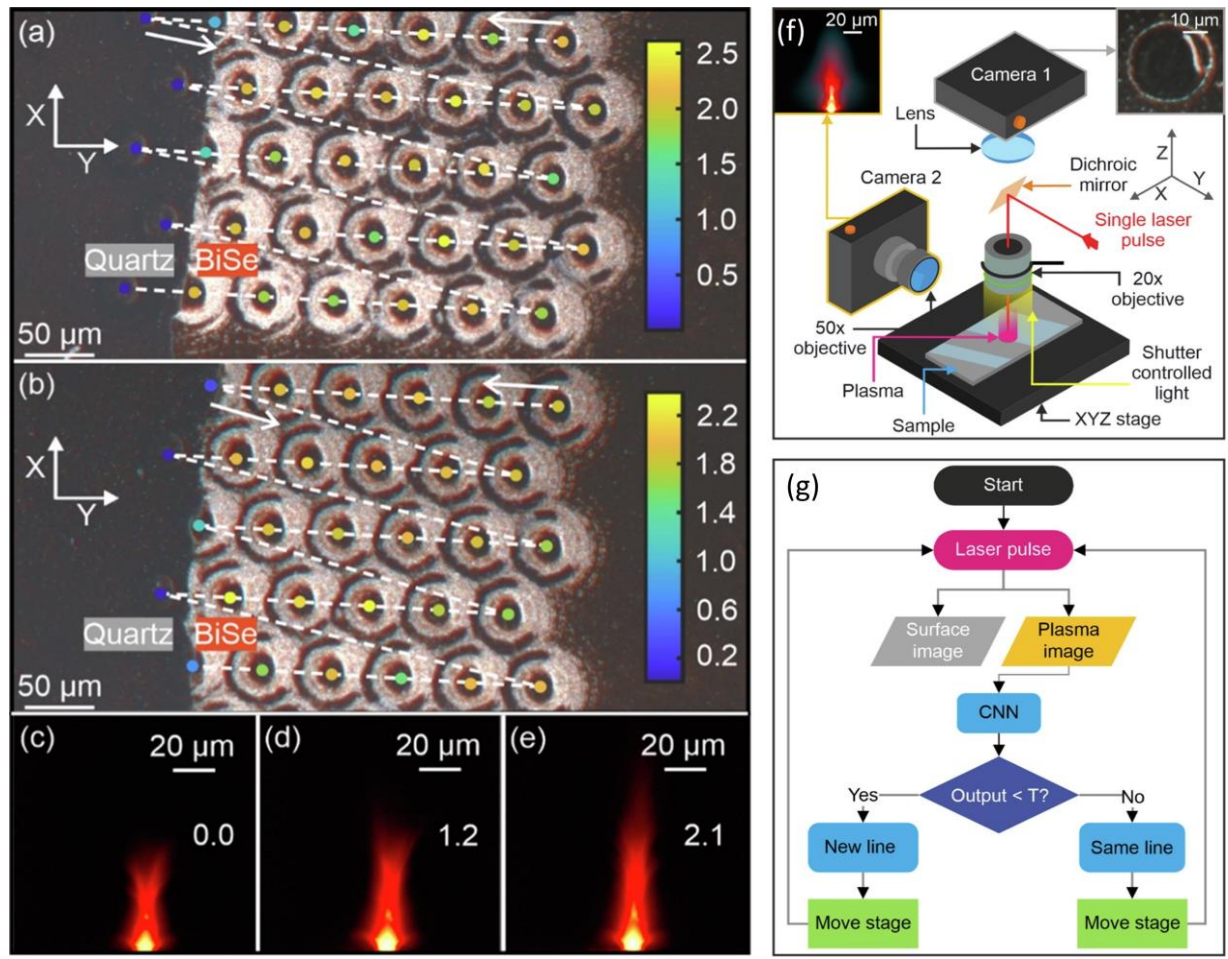

**Fig. 19.23:** Experimentally collected microscope images of laser ablation of uncoated-quartz and BiSe-coated-quartz, with the corresponding regression output values for the plasma images taken at each position displayed as solid circles in the figure, for threshold. f) schematic of the experimental setup and g) flow diagram of the applied strategy. Adapted from [72], Grant-Jacob, J.A., Mills, B. & Zervas, M.N. (2023) Real-time control of laser materials processing using deep learning. Manufacturing Letters, 38, 11–14. Copyright 2023 under Creative Commons BY 4.0 license. Retrieved from https://doi.org/10.1016/j.mfglet.2023.08.145.

The experiments successfully demonstrated the following:

1. *Boundary Detection:* The system effectively halted the laser focus when transitions occurred between materials, such as air and silica, or between uncoated quartz and BiSe-coated quartz.
2. *Automated Translation*: The stages were translated in real time to new scan positions upon boundary detection, thereby enhancing machining precision.
3. *Adaptability:* The neural network demonstrated versatility, adapting rapidly to new material configurations after training on small datasets.
4. 

**Process Control by Acoustic Imaging**

In another study by Grant-Jacob et al., both plasma images and acoustic signals were utilized to monitor the laser ablation process [68]. Acoustic data was collected using a microphone placed approximately 2 millimetres beneath the sample, positioned outside the laser path to prevent damage. Each recording lasted 3 seconds, starting 1 second before the laser pulse was triggered, ensuring the capture of both the laser-induced signal and background noise for contextual analysis. To process the acoustic data, peak detection algorithms isolated the laser-induced signal, which was cropped to a 42.3 ms window centred on the peak. The signal was then transformed into the frequency domain using MATLAB's Audio Toolbox, converting it into spectral components. The resulting spectral data, filtered to retain 94 frequency bands (11.4–22.1 kHz) for improved signal quality, was resized into a 512 × 512 pixel RGB image format for compatibility with the neural network. The concept is shown in **Fig. 19.24**.

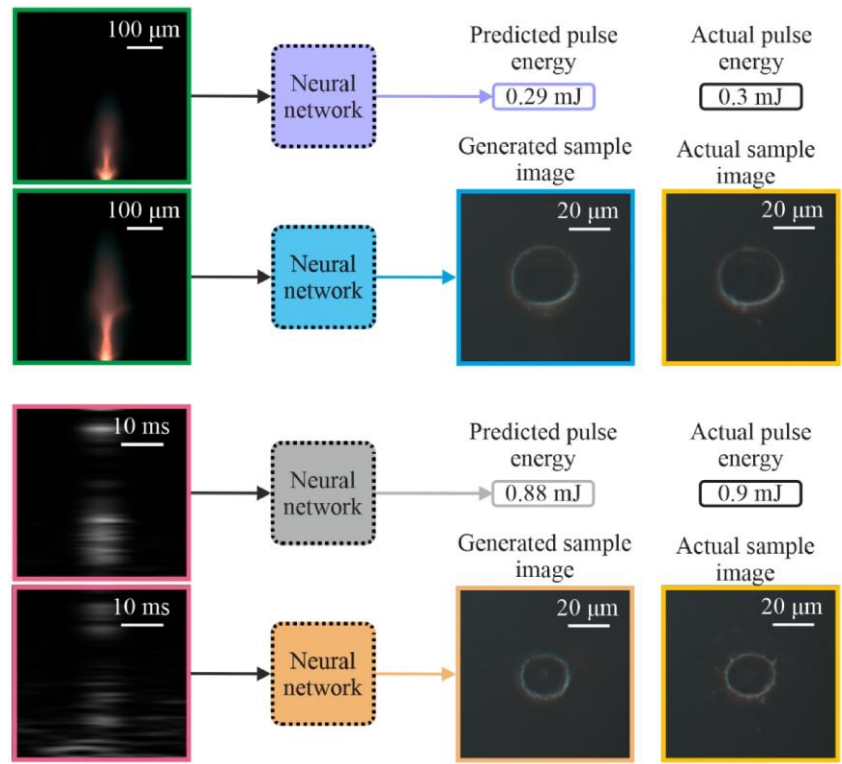

**Fig. 19.24:** Concept diagram of the application of the four neural networks used in this work, showing the use of plasma images and acoustic spectra for predicting the laser pulse energy and for predictive visualization of the appearance of the laser ablated samples. Reprinted from [68] , Grant-Jacob, J.A., Mills, B. & Zervas, M.N. (2023) Acoustic and plasma sensing of laser ablation via deep learning. Optics Express, 31(17), 28413–28422. Copyright 2023 under Creative Commons BY 4.0 license. Retrieved from https://doi.org/10.1364/OE.494700.

The results revealed that plasma-based monitoring provided superior accuracy compared to acoustic data. Plasma images achieved an R² value of 0.997, a standard deviation of 0.13 mJ, and a root-mean-square error (RMSE) of 0.02 mJ for pulse energy prediction. In contrast, predictions based on acoustic spectra were less precise, with an R² value of 0.957, a standard deviation of 0.30 mJ, and an RMSE of 0.05 mJ. Similarly, plasma images demonstrated greater fidelity in reconstructing ablated surface features, achieving a structural similarity index (SSIM) of up to 0.94, compared to 0.90 for acoustic-based predictions.The study concluded that plasma imaging was more effective for both pulse energy estimation and surface reconstruction due to the distinct and informative visual features of the plasma plume. Nonetheless, acoustic monitoring proved to be a valuable alternative in scenarios where direct visual observation was restricted, such as when plasma emissions obscure the workpiece.

# 4  Conclusion and Outlook

The application of machine learning in Laser Surface Texturing (LST) has proven transformative, advancing the precision, efficiency, and versatility of laser machining processes. Algorithms such as GA-ANFIS, ANN (with AutoDAQ), and GWO-BPNN are at the forefront, enabling the prediction and optimization of complex laser-textured surfaces with high accuracy. These methods enhance the quality control, minimize the need for trial-and-error experimentation, and offer insights into surface properties such as roughness and functionality. An overview of the presented algorithms adopted in this review article are presented in the following table with respect of their advantages in disadvantages for applicability and examples used for LST (**Tab. 19.1**).

Tab.19.1: Overview of the presented algorithms adopted in the review article.

| Algorithm | Advantages | Disadvantages | Examples |
|---|---|---|---|
| **GA-ANFIS** | Hybridization improved prediction and optimization accuracy. Effective for discrete parameter spaces. | Requires significant computational resources and time for tuning genetic algorithms and fuzzy systems. | Ji et al. [20] |
| **ANN (with AutoDAQ)** | Scalable for large datasets, adaptable to varying materials, and robust for quality control. | Performance is highly dependent on the quality and quantity of training data; can be resource-intensive. | Moles et al. [16] Ji et al. [20] Petit et al [19] |
| **GWO-BPNN** | Enhanced prediction accuracy ($R^2 > 0.9$) for multiple outputs. Reliable for complex relationships. | May struggle with overfitting if not carefully tuned and requires significant computational effort. | Liu et al. [22] |
| **Random Forest (RF)** | Effective in handling both regression and classification tasks, robust against overfitting, interpretable | Computationally expensive for large datasets and high-dimensional feature spaces. | Steege et al.[34], Moles et al. [16] |
| **XGBoost** | High accuracy, efficient in handling large datasets, and strong performance on unstructured data. | Sensitive to noisy data and can overfit with too many boosting rounds. | Petit et al. [19] |
| **Deep Learning (DNN, CNN, GAN)** | Ability to model complex, non-linear relationships, excellent for large datasets, effective for predictive visualization. | Requires large datasets for training and substantial computational power, may suffer from overfitting if not tuned well. | Mills et al.[27], Heath et al.[25], Tani & Kobayashi [24] |
| **Support Vector Machines (SVM)** | Effective for small to medium-sized datasets, good for high-dimensional data. | Computationally intensive, especially with large datasets or high-dimensional spaces. | Petit et al. [19] |

However, challenges remain, particularly with the computational resources required for deep learning models and the dependency on large, high-quality datasets for training. While techniques like Random Forests and XGBoost provide robustness against overfitting and good interpretability, their computational demands can still limit real-time applications. In contrast, deep learning approaches such as Convolutional Neural Networks (CNNs) and Generative Adversarial Networks (GANs) offer predictive visualization and advanced process control capabilities but require careful tuning to avoid overfitting and resource inefficiency.

Looking forward, machine learning's role in laser surface processing is poised for further advancements, particularly in real-time process control, adaptive parameter optimization, and predictive maintenance. With the increasing accessibility of computational power and sophisticated algorithms, AI-driven laser processing systems are expected to achieve greater efficiency, reducing costs, and expanding the range of applications. The convergence of machine learning, real-time monitoring, and automation will continue to drive innovation in industries such as manufacturing, healthcare, and materials science, transforming laser processing into a smarter, more adaptable technology.

**Acknowledgements**

This work was partially financed by the SYNTECS project under grant agreement 101091514 as part of the HORIZON-CL4-2022-TWIN-TRANSITION-01-02 call.